\pdfoutput=1

\documentclass[11pt,twoside,a4paper,cmspaper,final,collab]{cms-tdr}
\def\svnVersion{441392P}\def\cmsCernNoTag{CERN-EP-2017-130}\def\cmsCernDate{\today}\def\cmsMessage{Published in the Journal of High Energy Physics as \href{http://dx.doi.org/10.1007/JHEP12(2017)142}{\doi{10.1007/JHEP12(2017)142.}}}

\begin{document}\cmsNoteHeader{SUS-16-047}

\hyphenation{had-ron-i-za-tion}
\hyphenation{cal-or-i-me-ter}
\hyphenation{de-vices}
\RCS$Revision: 432827 $
\RCS$HeadURL: svn+ssh://svn.cern.ch/reps/tdr2/papers/SUS-16-047/trunk/SUS-16-047.tex $
\RCS$Id: SUS-16-047.tex 432827 2017-11-06 15:53:42Z kiesel $
\newlength\cmsFigWidth
\ifthenelse{\boolean{cms@external}}{\setlength\cmsFigWidth{0.85\columnwidth}}{\setlength\cmsFigWidth{0.4\textwidth}}
\ifthenelse{\boolean{cms@external}}{\providecommand{\cmsLeft}{top\xspace}}{\providecommand{\cmsLeft}{left\xspace}}
\ifthenelse{\boolean{cms@external}}{\providecommand{\cmsRight}{bottom\xspace}}{\providecommand{\cmsRight}{right\xspace}}

\newcommand{\EMHT}{\ensuremath{H_\mathrm{T}^{\gamma}}\xspace}
\newcommand{\EMHThlt}{\ensuremath{H_\mathrm{T}^{\gamma\text{,HLT}}}\xspace}
\newcommand{\gamjet}{\ensuremath{\gamma}+jet\xspace}
\newcommand{\gaugino}{\ensuremath{\widetilde{\chi}_1^{0/\pm}}\xspace}
\newcommand{\neutralino}{\PSGczDo}
\newcommand{\chargino}{\PSGcpmDo}
\newcommand{\gra}{\PXXSG}
\newcommand{\reclumi}{35.9\fbinv}

\cmsNoteHeader{SUS-16-047}

\title{Search for supersymmetry in events with at least one photon, missing transverse momentum, and large transverse event activity in proton--proton collisions at $\sqrt{s}=13\TeV$}

\date{\today}

\abstract{A search for physics beyond the standard model in final states with at least one photon, large transverse momentum imbalance, and large total transverse event activity is presented. Such topologies can be produced in gauge-mediated supersymmetry models in which pair-produced gluinos or squarks decay to photons and gravitinos via short-lived neutralinos. The data sample corresponds to an integrated luminosity of 35.9\fbinv of proton--proton collisions at $\sqrt{s}=13\TeV$ recorded by the CMS experiment at the LHC in 2016. No significant excess of events above the expected standard model background is observed. The data are interpreted in simplified models of gluino and squark pair production, in which gluinos or squarks decay via neutralinos to photons. Gluino masses of up to 1.50--2.00\TeV and squark masses up to 1.30--1.65\TeV are excluded at 95\% confidence level, depending on the neutralino mass and branching fraction.}

\hypersetup{
  pdfauthor={CMS Collaboration},
  pdftitle    = {Search for supersymmetry in events with at least one photon, missing transverse momentum, and large transverse event activity in proton-proton collisions at sqrt(s) = 13 TeV},
  pdfsubject  = {CMS SUSY analysis using photons and MET},
  pdfkeywords = {CMS, SUSY, photons, GMSB},
}

\maketitle

\section{Introduction} \label{sec:intro}

The standard model (SM) of particle physics describes elementary particles and
their interactions successfully.
Nevertheless, fine tuning of fundamental physics parameters is needed to cancel large quantum corrections
to the mass term in the Higgs potential~\cite{Barbieri198863}.
This and other problems of the SM can be addressed by supersymmetry
(SUSY) models~\cite{Ramond,Golfand:1971iw,Ferrara:1974pu,Wess,Chamseddine,Barbieri,Hall},
in which a SUSY partner particle is predicted for each SM particle.
Gauge-mediated SUSY breaking (GMSB) models~\cite{GGMa,GGMd2,GGMd3,GGMd4,GGMd5,GGMd1,GGMd}
allow for a natural suppression of flavour
violations in the SUSY sector
and can give rise to final states with photons and jets~\cite{Grajek:2013ola}.

The conservation of $R$ parity~\cite{rparity,Barbier:2004ez} implies that SUSY particles are
produced in pairs and the lightest SUSY particle (LSP) is stable.
If the LSP is neutral and only weakly interacting, it can escape detection,
leading to an imbalance of the total observed
transverse momentum. In this analysis, $R$-parity conservation is assumed
and the LSP is considered to be a nearly massless gravitino \gra.
The next-to-lightest-supersymmetric particle is assumed to be a gaugino \gaugino, which is a mixture of the
superpartners of the electroweak gauge bosons and the Higgs bosons. It decays promptly to a SM boson and a gravitino.
Both bino- and wino-like neutralinos \neutralino can decay to a photon and a gravitino;
wino-like charginos \chargino decay typically to a \PW~boson and a gravitino~\cite{Ruderman:2011vv}.
In this analysis, we assume gauginos are produced in decay chains of primary squarks or gluinos,
so the events also contain jets and thus large transverse event activity.

In this paper, a search for physics beyond the standard model (BSM) in final
states with at least one photon, large missing transverse momentum, and large
total transverse event activity is reported. The data used in this analysis were collected with the
CMS detector at the CERN LHC in 2016, and correspond to an integrated
luminosity of \reclumi of proton--proton collisions at a centre-of-mass energy
$\sqrt{s}=13\TeV$. Similar searches yielding no evidence for BSM physics have been performed at lower
centre-of-mass energies by CMS~\cite{CMS-PAPERS-SUS-14-004} with similar and alternative
selections~\cite{Khachatryan:2016hns,Khachatryan:2016ojf} and by the ATLAS
Collaboration~\cite{Aad:2015hea,ATLASCollaboration:2016wlb}.
The higher $\sqrt{s}$ of this dataset allows us to extend the
sensitivity to more massive SUSY particles.

\section{The CMS detector} \label{sec:cms}

The central feature of the CMS apparatus is a superconducting solenoid of
6\unit{m} internal diameter, providing a magnetic field of 3.8\unit{T}.
Within the solenoid volume are a silicon pixel and strip tracker, a lead
tungstate crystal electromagnetic calorimeter (ECAL),
and a brass and scintillator hadron calorimeter (HCAL),
each composed of a barrel and two endcap sections.
The electromagnetic calorimeter consists of 75\,848 lead tungstate crystals,
which provide coverage in pseudorapidity $\abs{\eta} < 1.48 $ in a barrel region (EB)
and $1.48 < \abs{\eta} < 3.0$ in two endcap regions (EE).
Forward calorimeters extend the pseudorapidity coverage provided by
the barrel and endcap detectors.
Muons are measured in gas-ionization detectors embedded in the steel
flux-return yoke outside the solenoid.
The jet energy resolution amounts typically to 15, 8, and 4\% at 10, 100, and 1000\GeV, respectively,
when combining information from the entire detector~\cite{Khachatryan:2016kdb}.
A more detailed description of the CMS detector, together with a definition of
the coordinate system used and the relevant kinematic variables,
can be found in Ref.~\cite{Chatrchyan:2008zzk}.

\section{Event reconstruction} \label{sec:rec}

The particle-flow (PF) algorithm reconstructs and identifies each individual
particle with an optimized combination of information from the various
elements of the CMS detector~\cite{Sirunyan:2017ulk}.
The energy of photons is directly obtained from the ECAL measurement.
The energy of electrons is determined from a combination of the electron
momentum at the primary interaction vertex as measured by the tracker,
the energy of the corresponding ECAL cluster,
and the energy sum of all bremsstrahlung photons spatially compatible with
originating from the electron track.
The momentum of muons is obtained from the curvature of the corresponding track.
The energy of charged hadrons is determined from a combination of their momentum
measured in the tracker and the matching ECAL and HCAL energy deposits,
corrected for zero-suppression effects and for the response function of the
calorimeters to hadronic showers.
Finally, the energy of neutral hadrons is obtained from the corresponding
corrected ECAL and HCAL energies.

Loose quality criteria with a selection efficiency close to 90\% are applied to
photons, based on the shower shape width in $\eta$, the hadronic energy fraction,
and the isolation from other particles. To distinguish photons from electrons,
photon candidates are not allowed to be associated with pixel seeds. Pixel seeds consist of two or
three hits in the pixel detector matching to the hypothetical trajectory
from the proton--proton interaction point to the energy cluster in the ECAL,
taking into account positively and negatively charged electron hypotheses.

Jets are reconstructed from all PF candidates, clustered by the
anti-\kt algorithm~\cite{Cacciari:2008gp, Cacciari:2011ma} with a
distance parameter of 0.4. To reduce the effect of additional proton--proton
collisions from the same or adjacent beam crossing (pileup) other than the primary
hard scattering process, charged hadrons from vertices not being
the primary vertex are excluded. An offset correction is applied to jet
energies to take the contribution from pileup interactions into
account~\cite{Cacciari:2007fd}.
The jet momentum is determined as the vector sum of momenta of all PF
candidates clustered into the jet. To correct for this, jet energy corrections
are applied, derived from
simulation and data using multijet, \gamjet, and leptonic Z+jets events.

The missing transverse momentum~\ptvecmiss is defined as the negative vector
sum of the transverse momenta \pt of all PF candidates in the event,
and its magnitude is denoted by \ptmiss. In order to improve the momentum resolution,
the jet energy corrections are propagated to \ptmiss.
The total transverse momentum \EMHT is the scalar sum of all jet momenta and the
\pt of the leading photon. Only jets with $\pt>30\GeV$ and $\abs{\eta}<3$ are
considered. In addition, if a jet is found within $\DR<0.4$ from the leading
photon, it is assumed that the jet \pt originates from the photon and the
jet \pt is not included in the  calculation of \EMHT.

\section{Signal models and event simulation} \label{sec:sim}

Monte Carlo (MC) generated events are used to study the SM backgrounds,
develop and validate the background estimation techniques, and model signal scenarios.
To generate
\gamjet, multijet, \PZ, \PW, \ttbar, $\gamma\PW$, $\gamma\PZ$, gluino pair,
and squark pair events, the \MGvATNLO 2.2.2~\cite{Alwall:2014hca} generator is used
at leading-order (LO) accuracy, while the next-to-leading-order
(NLO) accuracy is used for $\gamma\ttbar$ events.
The NNPDF3.0~\cite{Ball:2014uwa} parton distribution functions (PDFs) are used
in conjunction with \PYTHIA8.205 or 8.212~\cite{Sjostrand:2007gs} with the CUETP8M1
generator tune~\cite{Khachatryan:2015pea} for simulating parton showering and hadronization.
The LO cross sections are used for \gamjet events and events comprising solely jets
produced through the strong interaction (multijet events).
For all other background processes, NLO cross sections are used.
The contribution of pileup events is added to the hard scattering process
such that the probability of pileup events to occur is the same as that in the
data, with on average approximately 23 interactions per bunch crossing.

\begin{figure}[tbh]
  \centering
  \includegraphics[width=.49\textwidth]{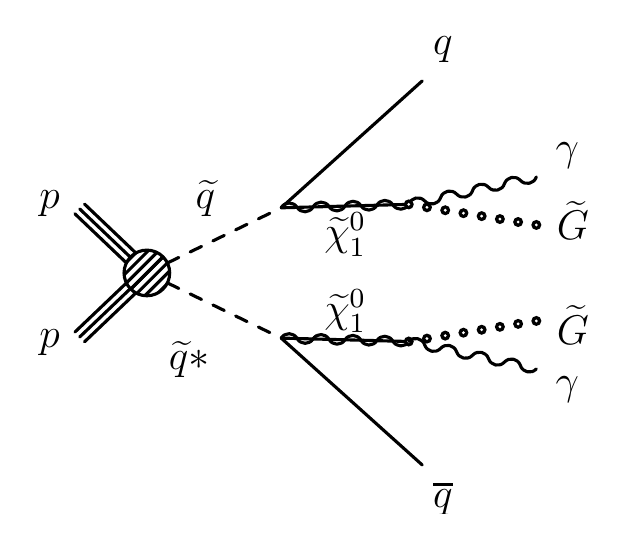}
  \includegraphics[width=.49\textwidth]{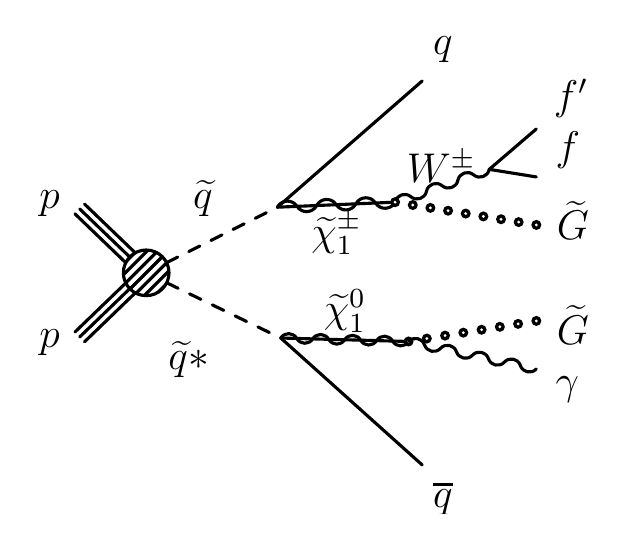}
  \includegraphics[width=.49\textwidth]{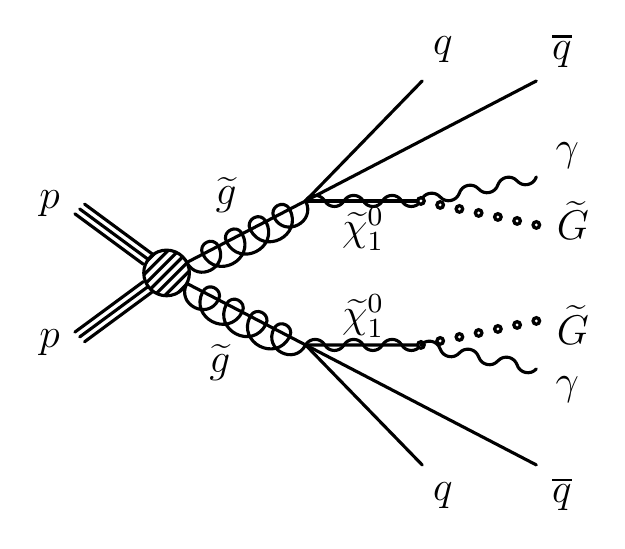}
  \includegraphics[width=.49\textwidth]{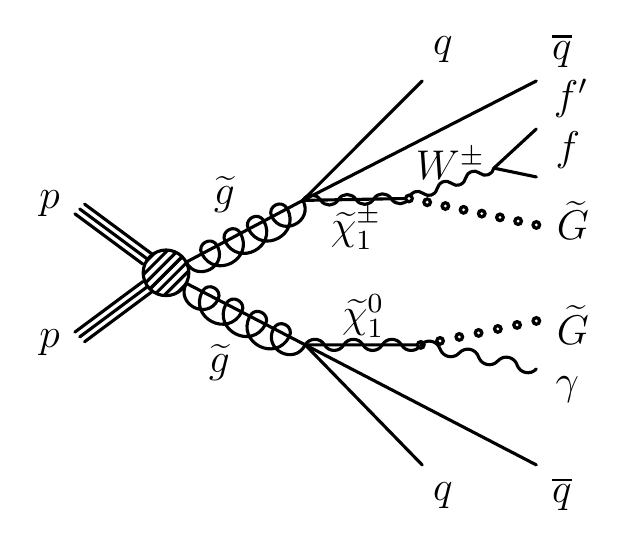}
  \caption{Feynman-like diagrams for the T6gg (top left) and the T5gg (bottom
    left) processes, and representative Feynman-like diagrams for the T6Wg (top
    right) and T5Wg (bottom right) processes. The T6Wg and T5Wg models include
    also diagrams with either two photons or two W bosons in the final state.
  }
  \label{fig:feynman}
\end{figure}

Gluino and squark pair production cross sections are determined using NLO plus
next-to-leading logarithm (NLL) calculations~\cite{Borschensky:2014cia}. Four
simplified models~\cite{Alves:2011wf,Chatrchyan:2013sza} are considered. The
T6gg model, where a first- or second-generation squark-antisquark pair is
produced, followed by the (anti)squark decay into an (anti)quark and a
neutralino. The neutralino decays promptly to a photon and a gravitino,
resulting in a final state with two jets, two photons, and missing transverse
momentum from the two gravitinos escaping detection. The T6Wg model is similar,
except the squarks decay with a probability of 50\% to a quark and a neutralino,
and a 50\% probability to decay to a quark and a chargino. The chargino further decays to a
\PW~boson and a gravitino, resulting in signatures with at least two jets, two
gravitinos, and two bosons. These two bosons can either be two photons, one
photon and one \PW~boson, or two \PW~bosons. The T5gg and T5Wg models consist of
gluino pair production. For these models, the squark masses are assumed to be much
larger than the gluino mass, leading to a three-body decay of the gluino to two
jets and a gaugino. For the T5gg model, the gauginos are neutralinos, while for
the T5Wg model, the gluino can also decay to jets and a chargino. Branching
fractions are assumed to be 100\%, except the squark to neutralino branching fraction in the
T6Wg model and the gluino to neutralino decay in the T5Wg model, which are 50\%
each. Feynman-like diagrams of these processes are shown in Fig.~\ref{fig:feynman}.

The CMS detector response is simulated using \GEANTfour~\cite{Agostinelli:2002hh} for SM processes,
while for signal events we use the CMS fast simulation~\cite{CMS-DP-2010-039,Sekmen:2017hzs}.
In the latter case, scale factors are applied to account for any differences with respect to the full simulation.
Event reconstruction is performed in the same manner as for collision data.

\section{Event selection and background prediction strategy} \label{sec:sel}

The high-level trigger system~\cite{Khachatryan:2016bia} selects events containing at least one photon
with $\pt>90\GeV$ and $\abs{\eta}<2.5$, and $\EMHThlt>600\GeV$,
where $\EMHThlt$ is defined as the scalar sum of the \pt for all jets passing the
kinematic selection used to select jets for the offline \EMHT calculation.
The trigger does not distinguish between jets and photons. As a result,
photons in the event, including the leading photon, are reconstructed as jets
and thus included in the calculation of $\EMHThlt$.
The efficiency for both the photon and the $\EMHThlt$ criterion are
measured independently, and their product is estimated to be equal to $(96\pm4)\%$, where
the uncertainty covers variations of the trigger efficiency versus time and versus
photon identification variables.

Events are selected if they contain at least one photon with $\pt>100\GeV$ in
the EB with $\abs{\eta}<1.4442$.
To reliably predict the background,
the photon is not allowed to be parallel or anti-parallel to \ptvecmiss within
an azimuthal angle of $\abs{\Delta\phi(\pm\ptvecmiss,\ptvec^\gamma)}<0.3$.
Three high-\ptmiss ranges (350--450, 450--600, and ${\geq}600\GeV$) and
two \EMHT selections (700--2000 and $\geq\!\!2000\GeV$) give rise to the definition
of six search regions. Additional selection
criteria are applied to remove events with spurious signals from instrumental
noise~\cite{CMS-PAS-JME-16-004}.
Background contributions of multijet, \gamjet, $\gamma\PZ$, $\gamma\PW$, $\gamma\ttbar$,
\PW+jets, and \ttbar events are estimated as described below.

\subsection{Background contribution of events with nongenuine \texorpdfstring{\ptmiss}{ptmiss}}
\label{subsec:fakeptmiss}

A small fraction of \gamjet events can populate the signal region because of
artificial \ptmiss generated by momentum mismeasurement in the detector.
Jets have the largest transverse momentum uncertainties, and even though the probability of a
large mismeasurement is low, the large cross section of the \gamjet process leads to
a nonnegligible contribution to the search region.
Multijet events have an even higher cross section, and contribute to the signal
selection if one of the jets is misidentified as a photon.
As in \gamjet events, nonzero \ptmiss in multijet events is caused by the finite
jet momentum resolution.

Estimating these backgrounds from simulation would result in a large uncertainty for two reasons:
the large cross section requires a large number of simulated events to obtain
a small statistical uncertainty; in addition, small differences between the
measured and simulated jet response can lead to large differences at high \ptmiss
values between measured and simulated events.
A background estimation method based on control samples in data was therefore
developed to achieve
smaller uncertainties without relying on the simulated jet energy response.
We performed this method independently for the low- and high-\EMHT selection.
The shapes of the \ptmiss distributions in \gamjet and multijet events are found
to be similar, and their normalizations can be extracted from low-\ptmiss events,
where no significant signal contribution should be present.
This is verified using simulated event samples.
We use the shape of the \ptmiss distribution of a multijet event sample as a prediction
for events with nongenuine \ptmiss.

For the background estimate, the photon control region (CR) is defined by
requiring the search selection, but requiring $\ptmiss<100\GeV$. A jet CR is
defined by selecting events with the \EMHT criteria only, based on a trigger
with only the \EMHThlt selection. For low \ptmiss values, the
jet CR is dominated by multijet events, but for large \ptmiss values,
$\PW(\ell\nu)$+jets, $\PZ(\nu\nu)$+jets, and \ttbar events can also contribute.
These are subtracted using simulation. The shape of the \ptmiss
distribution of \gamjet and multijet events in the photon CR is very similar to
that in the jet CR.

To correct for residual differences between the two CRs, a correction factor
is applied to the \ptmiss values of the jet CR. Studies showed that a constant
multiplicative factor leads to the best agreement between the \ptmiss shapes in
the two CRs. The factor is chosen such that it minimizes the
$\chi^2$ between the shapes of the \ptmiss distributions in the two CRs for
$\ptmiss<100\GeV$, and is about 0.90 (0.84) for the low- (high-) \EMHT selection.
The uncertainty in this factor is calculated as the quadratic sum
of the deviation of the factor from unity and the statistical uncertainty in the
$\chi^2$ method. The \ptmiss distribution of the jet CR is then scaled to the
\ptmiss distribution of the photon CR in $\ptmiss < 100\GeV$ to provide an
estimate for the background contribution of nongenuine \ptmiss events in the signal
selection. Several uncertainties are considered. The uncertainty associated to
the shift factor is obtained by multiplying the jet CR by the factor modulated by its
uncertainty. The uncertainty in the normalization is derived from the
statistical uncertainty of the photon CR and the jet CR in the
$\ptmiss<100\GeV$ range.  The statistical uncertainty assigned to the
prediction due to the number of events in the jet CR at high \ptmiss is about
as large as the systematic uncertainty.

\begin{figure}[tbh]
  \centering
  \includegraphics[width=.49\textwidth]{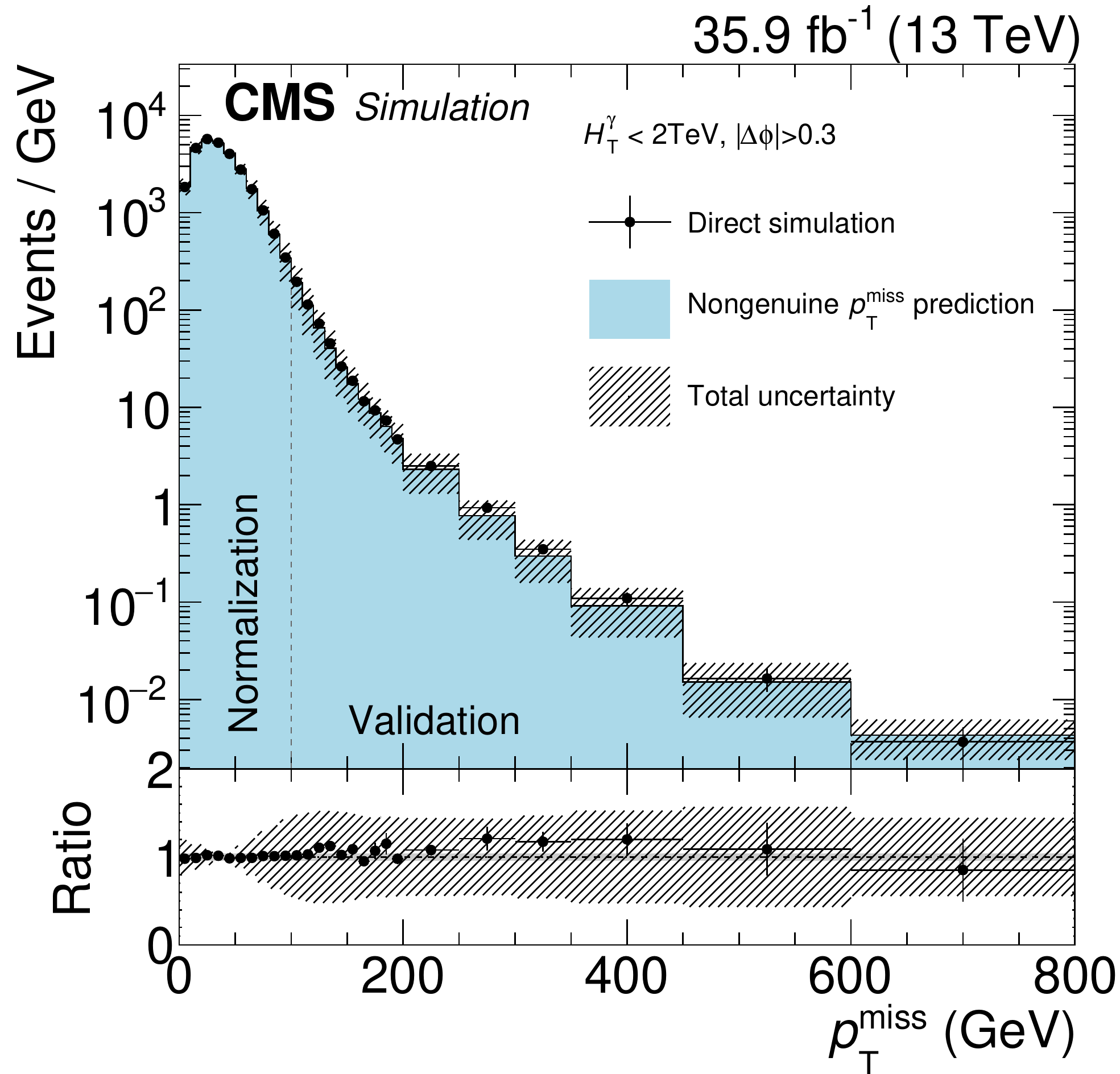}
  \includegraphics[width=.49\textwidth]{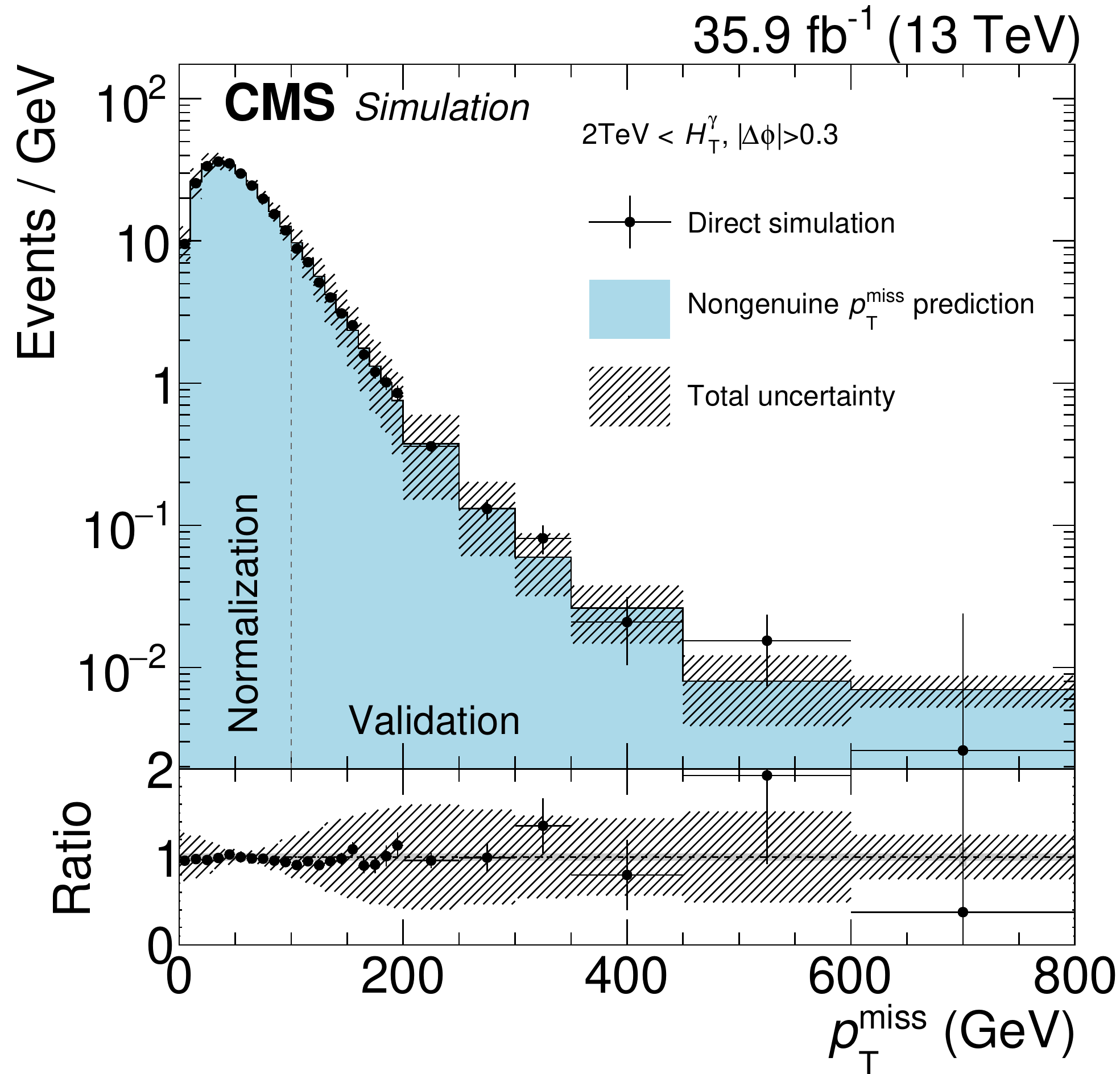}
  \caption{Validation of the nongenuine \ptmiss background estimation method with
    \gamjet and multijet simulations. The direct simulation results are shown as
    black dots, while the prediction using the jet CR is shown as light
    blue histogram. The total uncertainty of the prediction is presented as shaded
    area. The bottom panel shows the ratio of the direct simulation to the prediction.
    The low- (high-) \EMHT selection is shown on the left (right).
    The number of events corresponds to the expectation in data for an
    integrated luminosity of \reclumi.
    The rightmost bin includes all events with $\ptmiss>600\GeV$.
  }
  \label{fig:qcdClosure}
\end{figure}

The method is tested on simulated \gamjet and multijet events.
The comparison of direct simulation results and the prediction from simulation, using this method,
is shown in Fig.~\ref{fig:qcdClosure}.
In this figure and the following ones, the rightmost bin includes all events with $\ptmiss>600\GeV$.
The agreement between the two distributions suggests that the method is performing as expected.
Further validation is discussed in Section~\ref{subsec:validation}.

\subsection{Background contribution from events with electrons}

Electrons and photons have similar calorimetric response.
If no pixel seeds are reconstructed for an electron, it can be misidentified
as a photon.
In \PW+jets or \ttbar processes, electrons are produced in association with neutrinos,
so these events tend to also have large \ptmiss and enter the search regions.
To estimate the contribution of these processes, a CR with electrons
is defined and scaled by the electron-to-photon ($\Pe\to\gamma$) misreconstruction
probability.

The electron CR is defined similar to the search selection, except that the
photon candidate is required to have pixel seeds, thereby selecting events with
electrons. For high \ptmiss, this CR is dominated by W and \ttbar
events.

The electron-to-photon misreconstruction probability is estimated with the
tag-and-probe meth\-od using an event sample dominated by $\PZ\to\Pe\Pe$
events, and is 2.7\% for data and 1.5\% for simulation. For the prediction in
data, the probability measured with data is used, while for the validation in
simulation, the probability measured with simulated events is used. To account
for differences between the misreconstruction rate determined from the \PZ~boson resonance and the
\PW~boson dominated electron CR with high \ptmiss and high \EMHT,
a systematic uncertainty of 30\% is applied to the misreconstruction rate.
The size of the uncertainty is based on studies of the variation of the
misreconstruction probability versus various kinematic and geometric quantities
in data and simulation.

\begin{figure}[tbh]
  \centering
  \includegraphics[width=.49\textwidth]{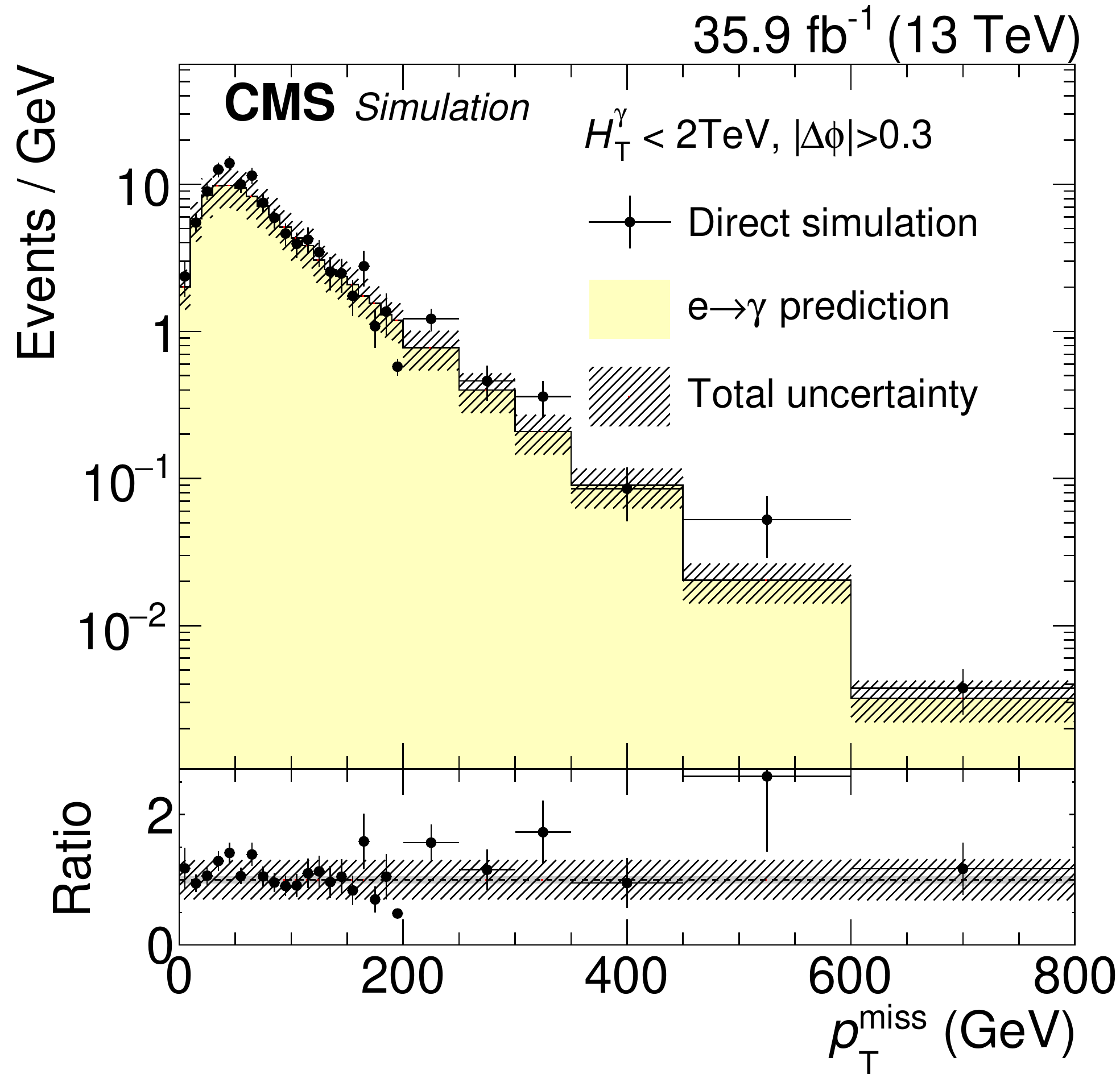}
  \includegraphics[width=.49\textwidth]{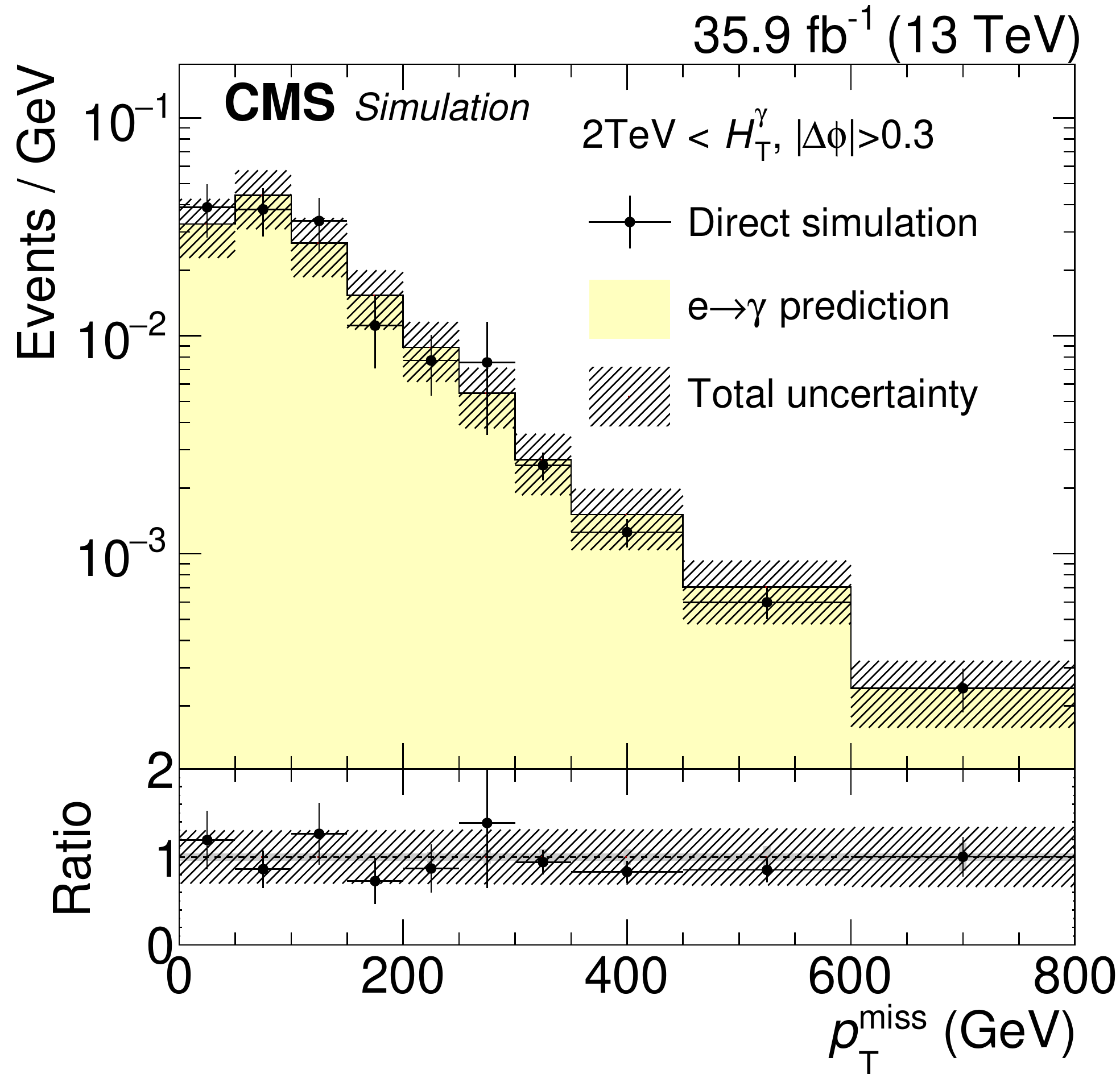}
  \caption{Validation of the background estimation method for electrons misreconstructed as photons
  using \PW+jets and \ttbar simulation.
  The low- (high-) \EMHT selection is shown on the left (right).
  The number of events corresponds to the expectation in data for an
  integrated luminosity of \reclumi.
    The rightmost bin includes all events with $\ptmiss>600\GeV$.
}
  \label{fig:electronClosure}
\end{figure}

The background estimation method is tested on simulated \PW+jets and \ttbar events.
The direct simulation of electrons reconstructed as photons is compared to the electron CR,
scaled by the electron-to-photon misreconstruction probability as shown in
Fig.~\ref{fig:electronClosure}, but including also low \ptmiss events.
The agreement in the search regions suggests that the method is performing as expected.

\subsection{Backgrounds estimated from simulation}\label{subsec:bgPredictionMC}

Also contributing to the search region are the processes $\gamma\PW(\ell\nu)$,
$\gamma\PZ(\nu\nu)$, and $\gamma$\ttbar, which are estimated using simulation.
Simulated events with electrons reconstructed as photons passing the event selection
are omitted since they are estimated
using data.
The photon in the event can be produced in the hard scattering or in the shower,
either as initial- (ISR) or final-state radiation, or as a jet misreconstructed
as a photon.
Events are simulated with and without a photon in the hard scattering process,
and the overlap between the samples is removed.
The reconstruction and identification efficiencies for photons are measured in
$\PZ\rightarrow\re\re$ and $\PZ\rightarrow\mu\mu\gamma$ data and simulation.
The ratio of these efficiencies is consistent with unity and has an uncertainty
of about 3\%. Simulated events are weighted by the ratio of the efficiencies, and the
uncertainty is propagated to the event yield.
The NLO cross sections are used, and several uncertainties are considered, with
their relative uncertainties given here in parentheses:
factorization and renormalization scales (16--27\%), PDFs (5--10\%)~\cite{Butterworth:2015oua},
contribution of pileup events (0.2--6\%), trigger efficiency (4\%), jet resolution and energy scales (2--20\%),
integrated luminosity (2.5\%)~\cite{CMS-PAS-LUM-17-001}, and statistical uncertainty of the simulated samples (4--47\%).
For the study of the renormalization and factorization scale uncertainties,
variations up and down by a factor of two with respect to the nominal values
of the scales are considered. The maximum difference in the yields with respect
to the nominal case is used as the uncertainty.
The pileup uncertainty corresponds to the variation of the number of predicted
events if the total inelastic proton--proton cross section is shifted by
$\pm$5\%.

\subsection{Validation of the background estimation methods}\label{subsec:validation}

\begin{figure}[tbh]
  \centering
  \includegraphics[width=.49\textwidth]{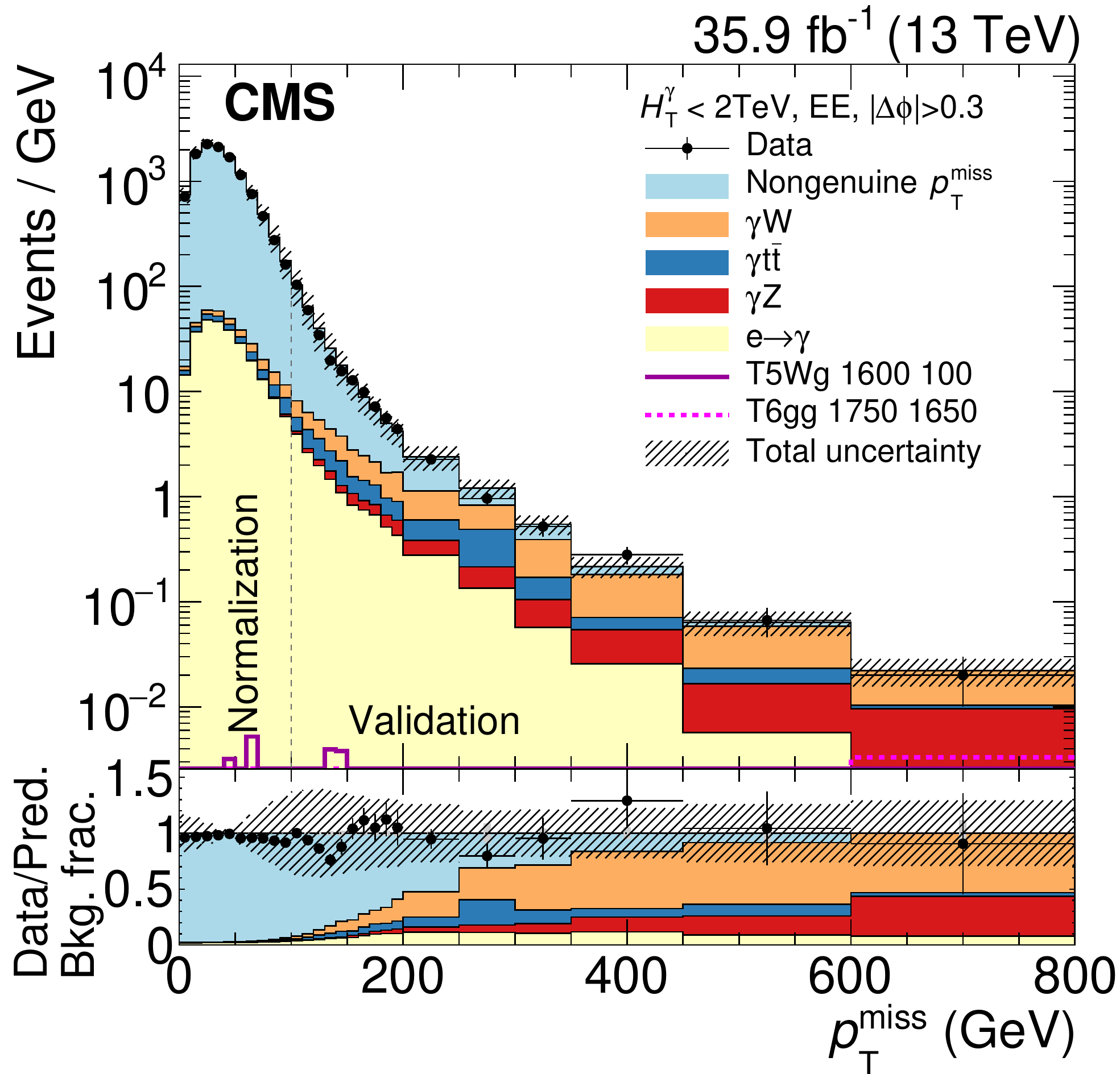}
  \includegraphics[width=.49\textwidth]{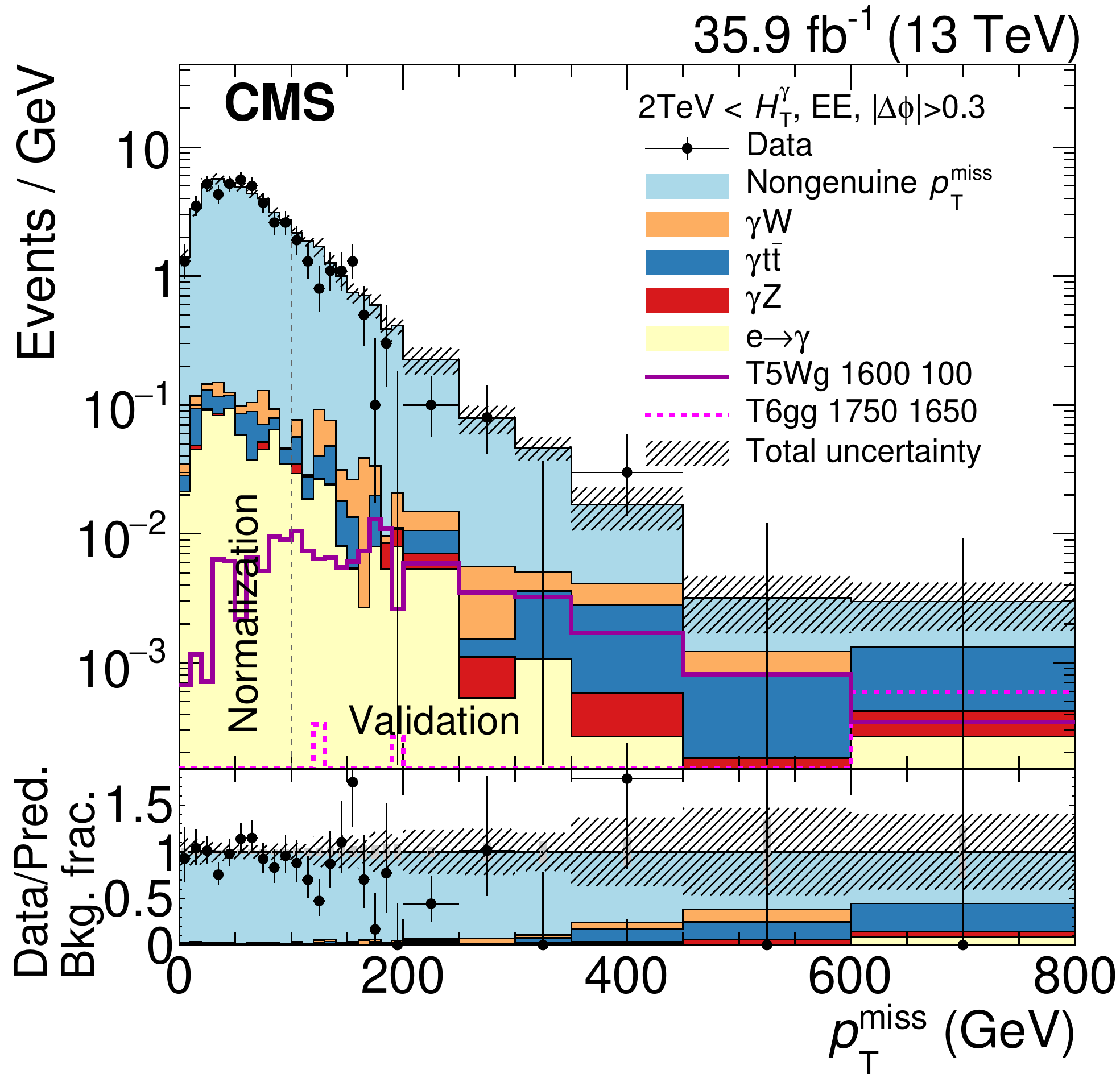}
  \caption{Validation of the background estimation methods with photons reconstructed
    in the EE. The expectation for the T5Wg signal scenario with a gluino mass of 1600\GeV
    and a gaugino mass of 100\GeV and the T6gg signal scenario with a squark mass
    of 1750\GeV and a neutralino mass of 1650\GeV are shown.
    The low- (high-) \EMHT selection is shown on the left (right).
    Below the \ptmiss distributions, the data divided by the background prediction
    are shown as black dots, and the relative background components are shown
    as coloured areas.
    The rightmost bin includes all events with $\ptmiss>600\GeV$.
    }
  \label{fig:val}
\end{figure}

In addition to the validation of the background estimation methods with simulated
events, the methods are also validated using data from two mutually exclusive event selections.
The first validation region is defined with noncentral photons.
Instead of the photon being reconstructed in the EB, the leading photon must be reconstructed
in the range $1.6<\abs{\eta}<2.5$.
This is not the full range of the EE, but in this range the background contribution
from electrons reconstructed as photons is similar to the one in the EB search
region.
High-mass gluinos and squarks tend to decay more centrally, leaving the EE validation
region essentially free of potential signal events.
The same methods as for the EB search regions are applied, and the resulting
distributions are shown in Fig.~\ref{fig:val}.
The \ptmiss distributions of two signal models are displayed as well.
In the low-\EMHT region and for large \ptmiss of the high-\EMHT region,
the observed number of events agrees with the prediction.
The second validation region is similar to the search regions with photons reconstructed
in the EB, with $100<\ptmiss<350\GeV$,
which is orthogonal to both the region used to normalize
the multijet background ($\ptmiss < 100\GeV$) as well as the signal
regions ($\ptmiss>350\GeV$), and is shown in Fig.~\ref{fig:final}.
Good agreement is observed in this validation region as well.

\section{Results} \label{sec:results}

\begin{figure}[tbh]
  \centering
  \includegraphics[width=.49\textwidth]{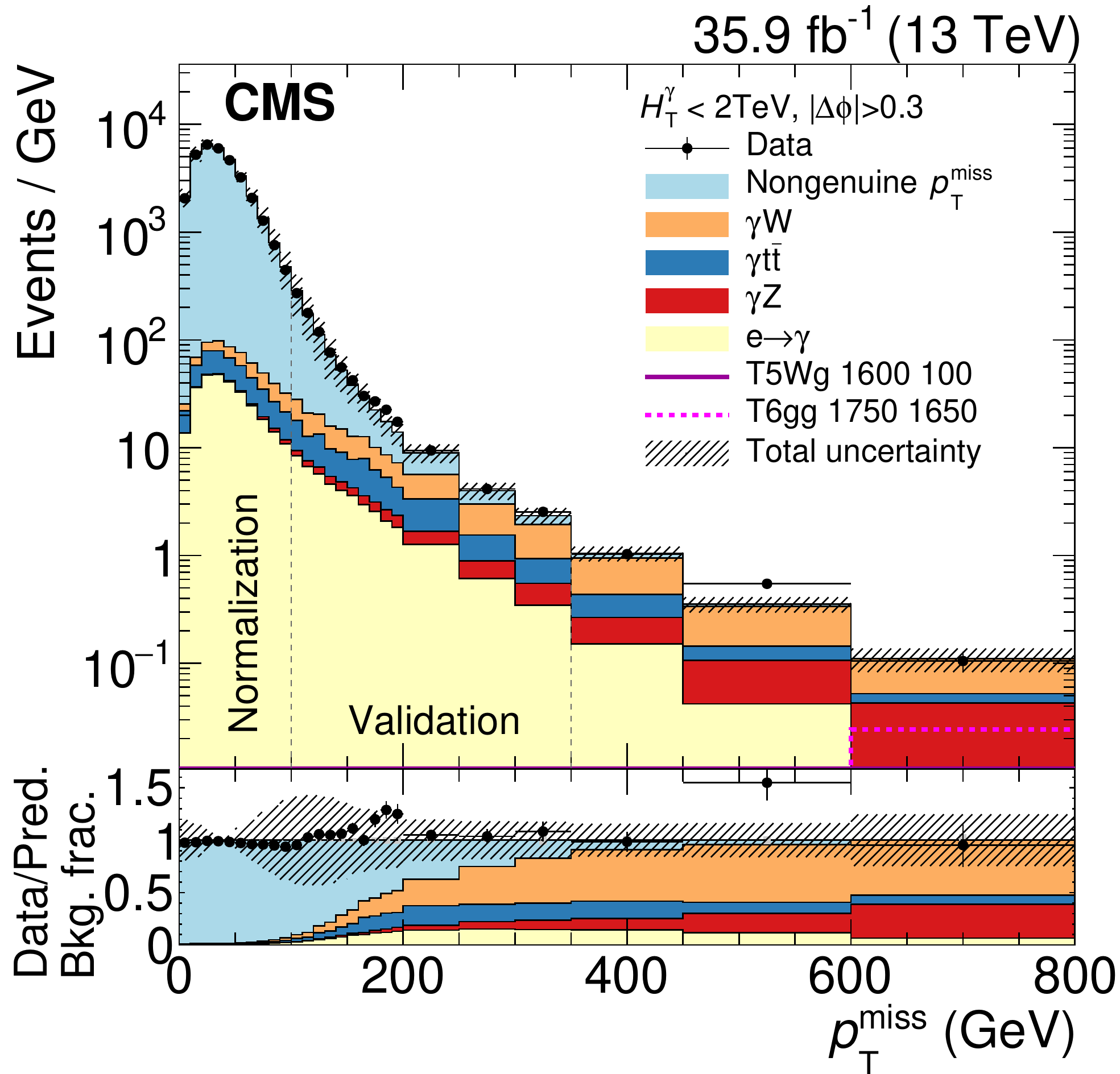}
  \includegraphics[width=.49\textwidth]{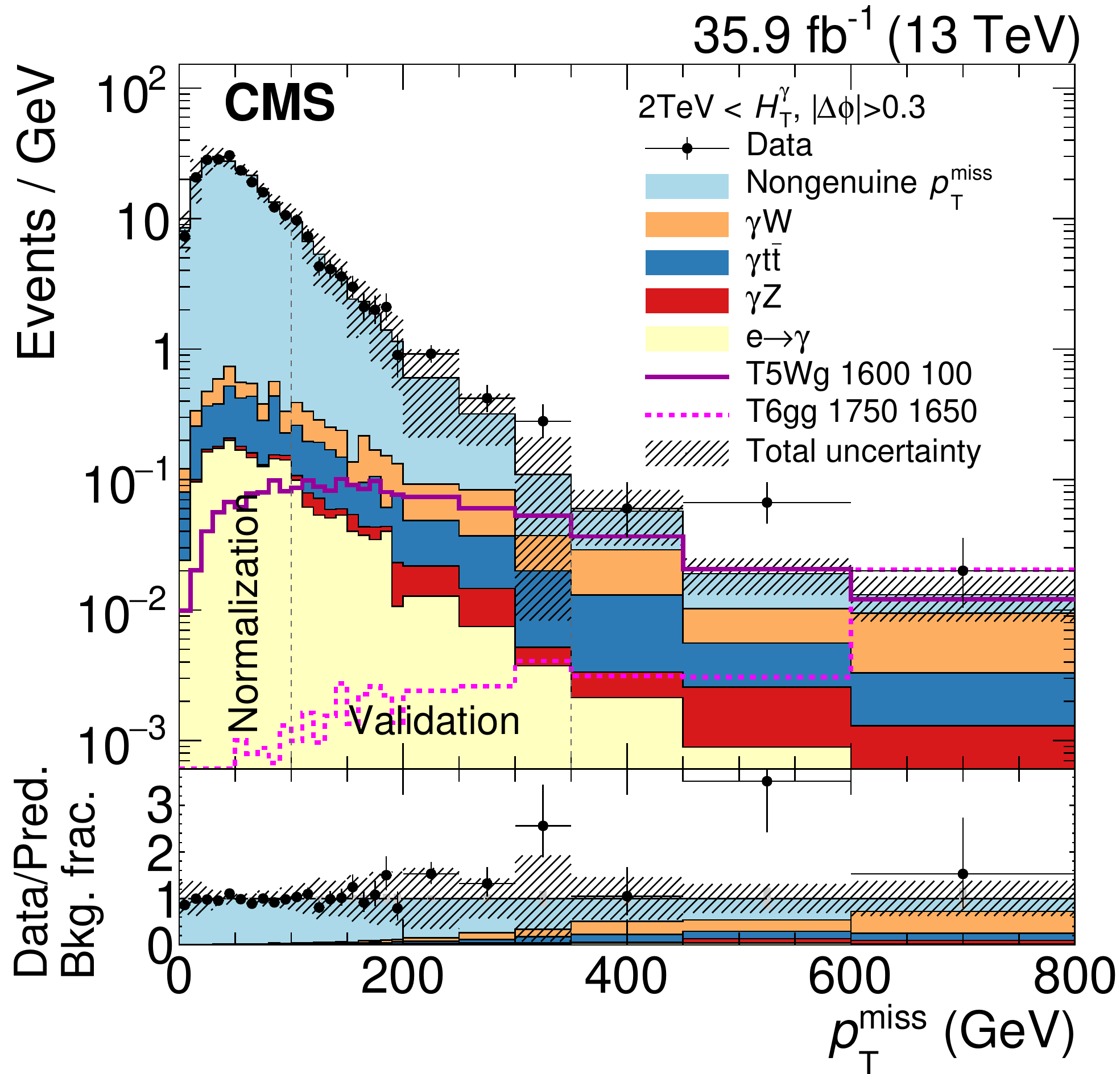}
  \caption{Observed data compared to the background prediction.
    The expectation for the T5Wg signal scenario with a gluino mass of 1600\GeV
    and a gaugino mass of 100\GeV and the T6gg signal scenario with a squark mass
    of 1750\GeV and a neutralino mass of 1650\GeV are shown.
    The low- (high-) \EMHT selection is shown on the left (right).
    Below the \ptmiss distributions, the data divided by the background prediction
    are shown as black dots, and the relative background components are shown
    as coloured areas.
    The last three bins in each plot correspond to the search regions.
    The rightmost bin includes all events with $\ptmiss>600\GeV$.
  }
  \label{fig:final}
\end{figure}

The predicted number of SM background events, the expected signal yield for two signal
scenarios and the number of observed events in data are shown in Fig.~\ref{fig:final} and Table~\ref{tab:final}.
The uncertainties (including the uncertainties for the signal models) are presented
in Table~\ref{tab:unc}.
The low-\EMHT search regions are dominated by $\gamma\PW$ events and are sensitive
to signal models with low squark or gluino masses.
The high-\EMHT search regions are dominated by background with nongenuine \ptmiss
and have larger sensitivity to models with high gluino or squark masses and low
gaugino masses.
Overall, the number of observed events is in agreement with the prediction.
The second search bin in both the low- and high-\EMHT regions shows an excess with
local significance of 1.9 and 2.7 standard deviations ($\sigma$), respectively.
In the highest \ptmiss bins, which are more sensitive for most signal scenarios,
the number of observed events is compatible with the background expectation.

\begin{table}[tbh]
  \topcaption{Observed data compared to the background prediction and the expected
    signal yields for two signal scenarios. The expectations are given for the T5Wg signal
    scenario with a gluino mass of 1600\GeV and a gaugino mass of 100\GeV and
    the T6gg signal scenario with a squark mass of 1750\GeV and a neutralino
    mass of 1650\GeV.
    The quadratic sum of statistical and systematical uncertainties is given.
    Only experimental uncertainties for the signal model are stated.}
  \label{tab:final}
\resizebox{\textwidth}{!}{
\begin{tabular}{l|rrr|rrr}
  \EMHT (\GeVns{}) & \multicolumn{3}{c|}{$<$2000} & \multicolumn{3}{c}{$>$2000}\\
  \ptmiss (\GeVns{}) & \multicolumn{1}{c}{(350, 450)} & \multicolumn{1}{c}{(450, 600)} & \multicolumn{1}{c|}{$>$600} & \multicolumn{1}{c}{(350, 450)} & \multicolumn{1}{c}{(450, 600)} & \multicolumn{1}{c}{$>$600} \\
  \hline
  Nongenuine \ptmiss & ${9.6\,\,}^{+\,\,\,\,11.1}_{-\,\,\,\,\phantom{0}9.6}\,$ &${2.2\,\,}^{+\,\,\,5.5}_{-\,\,\,2.2}\,$ &$<0.1$ &$2.83\pm2.51$ &$1.31\pm0.74$ &${0.73\,\,}^{+\,\,\,0.86}_{-\,\,\,0.73}\,\,$ \\
  $\gamma\PW$ & $51.3\pm\phantom{0}9.7$ &$29.1\pm5.5$ &$11.6\pm2.5$ &$1.58\pm0.58$ &$0.70\pm0.37$ &$1.23\pm0.43$ \\
  $\gamma\ttbar$ & $17.1\pm\phantom{0}5.4$ &$5.6\pm2.6$ &$1.9\pm0.4$ &$0.97\pm0.38$ &$0.45\pm0.29$ &$0.40\pm0.22$ \\
  $\gamma\PZ$ & $11.5\pm\phantom{0}2.4$ &$9.7\pm1.8$ &$7.1\pm1.4$ &$0.12\pm0.07$ &$0.25\pm0.11$ &$0.21\pm0.10$ \\
  $\Pe\to\gamma$ & $15.1\pm\phantom{0}4.6$ &$6.3\pm1.9$ &$1.4\pm0.5$ &$0.21\pm0.10$ &$0.13\pm0.07$ &$0.05\pm0.04$ \\
  Total bkg.& $104.6\pm16.5$ &$53.0\pm8.6$ &$22.0\pm3.0$ &$5.72\pm2.60$ &$2.84\pm0.89$ &$2.62\pm0.99$ \\
  \hline
  Data & $103\phantom{.0\pm 00.0}$ &$82\phantom{.0\pm 0.0}$ &$21\phantom{.0\pm 0.0}$ &$6\phantom{.00\pm 0.00}$ &$10\phantom{.00\pm 0.00}$ &$4\phantom{.00\pm 0.00}$ \\
  T5Wg \, 1600 \, 100 & $0.4\pm\phantom{0}0.1$ &$0.8\pm0.1$ &$0.7\pm0.1$ &$3.66\pm0.40$ &$3.09\pm0.40$ &$2.41\pm0.32$ \\
  T6gg \, 1750 \, 1650 &  $0.5\pm\phantom{0}0.1$ &$0.8\pm0.1$ &$4.9\pm0.4$ &$0.31\pm0.04$ &$0.46\pm0.07$ &$4.12\pm0.32$ \\
\end{tabular}
}
\end{table}

\section{Interpretation} \label{sec:interpretation}
The systematic uncertainties of the nongenuine \ptmiss background are fully
correlated within the high- and low-\EMHT selections, and are described in Section~\ref{subsec:fakeptmiss}.
The systematic uncertainty in the electron misidentification background is
fully correlated for all search regions, as are most uncertainties in the simulated
backgrounds described in Section~\ref{subsec:bgPredictionMC}.

To improve on the signal simulation of the multiplicity of additional jets
from ISR, simulated signal events are reweighted based on the number of ISR
jets ($N_\text{J}^\text{ISR}$) so as to make the jet multiplicity in simulated
\ttbar samples agree with that in data. The reweighting factors vary between
0.92 and 0.51 for $N_\text{J}^\text{ISR}$ between 1 and 6. We take one half of
the deviation from unity as the systematic uncertainty in these reweighting
factors, correlated between all search regions. The renormalization and
factorization scales, and PDF uncertainties in the cross sections for signal
simulation are taken from Ref.~\cite{Borschensky:2014cia}.
To estimate the influence of pileup in signal events, the selection is done
with a high and a low number of additional interactions. The difference in selection
efficiency is taken as a systematic uncertainty.
Since all physics objects are included in the computation of \ptmiss, it can be
difficult to describe accurately within the CMS fast simulation. The \ptmiss
of the models considered, however, is dominated by the missing momentum carried
away by the gravitons and not by the modelling of resolution effects. An
additional systematic uncertainty of between 0.5 and 6\% is assigned by
calculating the mean difference between the reconstructed and generated \ptmiss.
A summary of the uncertainties can be found in Table~\ref{tab:unc}.

\begin{table}[tbh]
  \centering
  \topcaption{Systematic uncertainties for background determined from control
  samples in data (first two rows)
  and simulation (all other rows). If two values are given, the first one is for
  simulated SM backgrounds, while the latter is for simulated signal.
  The PDF and scale uncertainties for the signal simulation affect the shape only, as
  the uncertainty in the rate is already considered in the overall cross section uncertainty~\cite{Borschensky:2014cia}.}
  \label{tab:unc}
\begin{tabular}{l|cc}
 & \multicolumn{2}{c}{Relative uncertainty (\%)} \\
Source & background & signal \\\hline
Nongenuine \ptmiss & 14--250 & \\
$\Pe\to\gamma$ & 30 & \\\hline
Integrated luminosity & 2.5 & 2.5 \\
Photon scale factors & 2 & 2 \\
Trigger & 4 & 4 \\
PDFs & 5--10 &  \\
Renormalization/factorization scales & 16--27 & 0--1\\
Jet energy scale and resolution & 2--20 & 1--6\\
Pileup & 0.2--6 & 0.2--10\\
ISR &  & 0--10\\
Fast simulation \ptmiss modelling &  & 0.5--6\\
\end{tabular}
\end{table}

\begin{figure}[tbh]
  \centering
  \includegraphics[width=.49\textwidth]{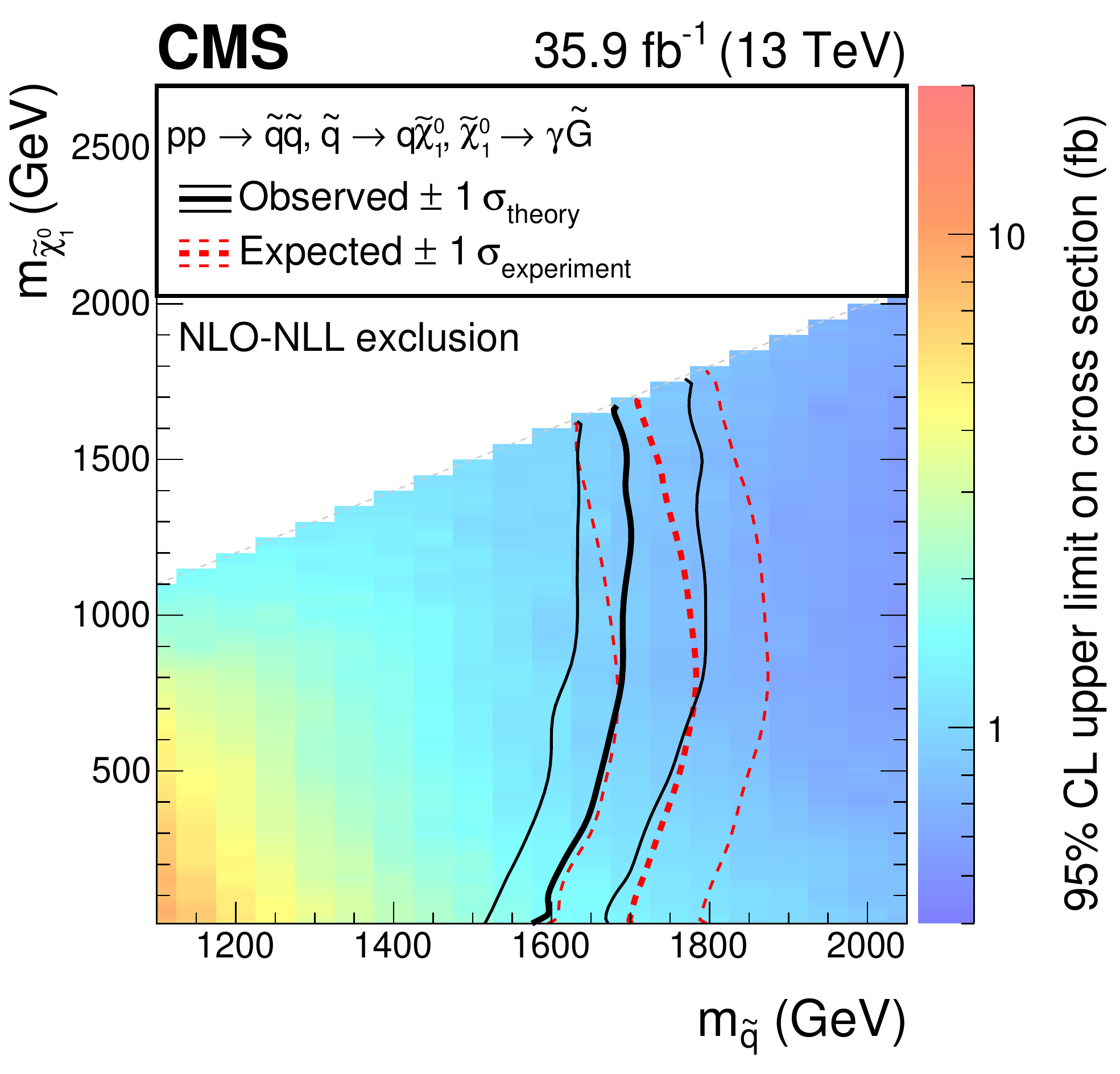}
  \includegraphics[width=.49\textwidth]{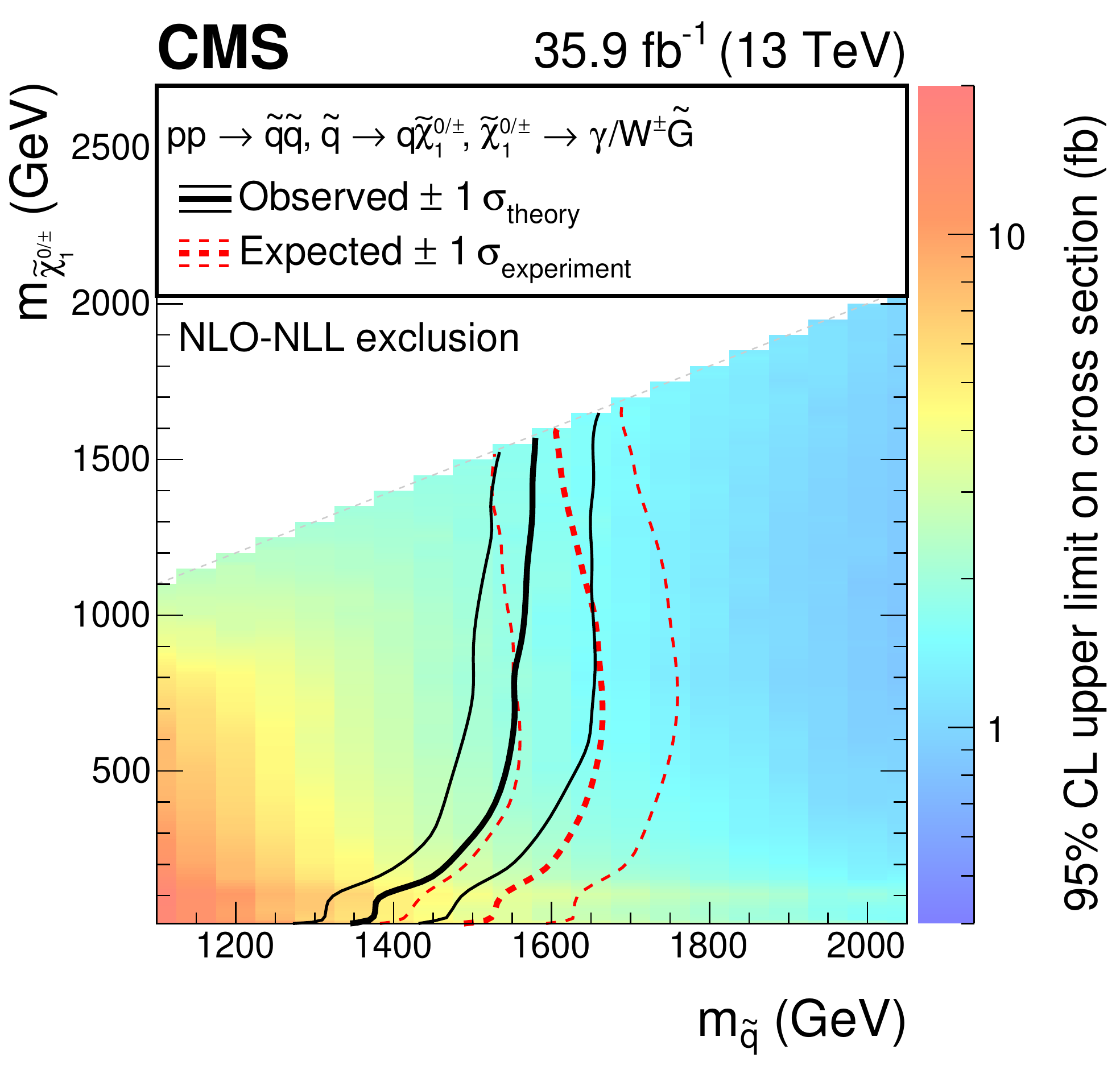}
  \includegraphics[width=.49\textwidth]{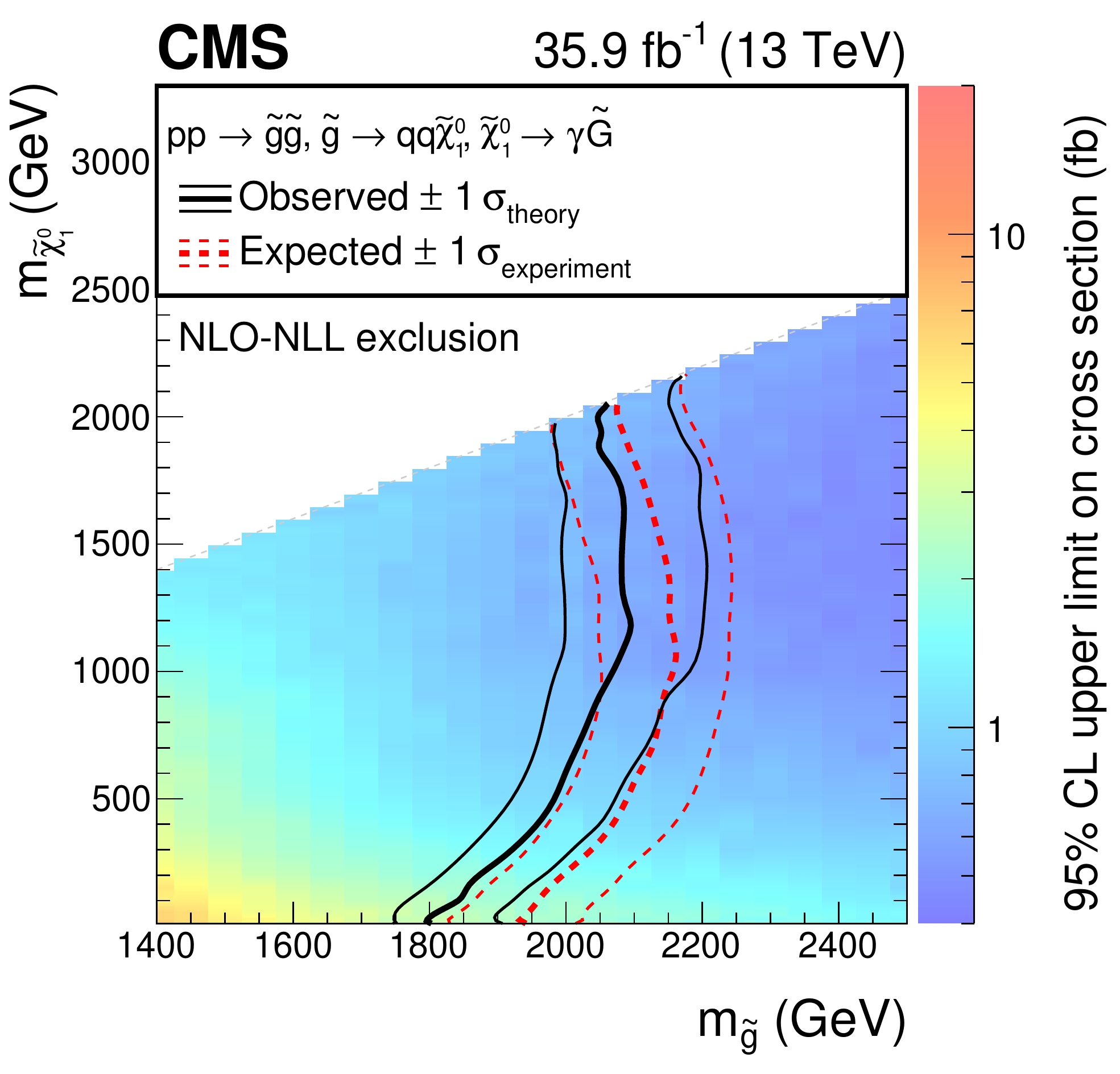}
  \includegraphics[width=.49\textwidth]{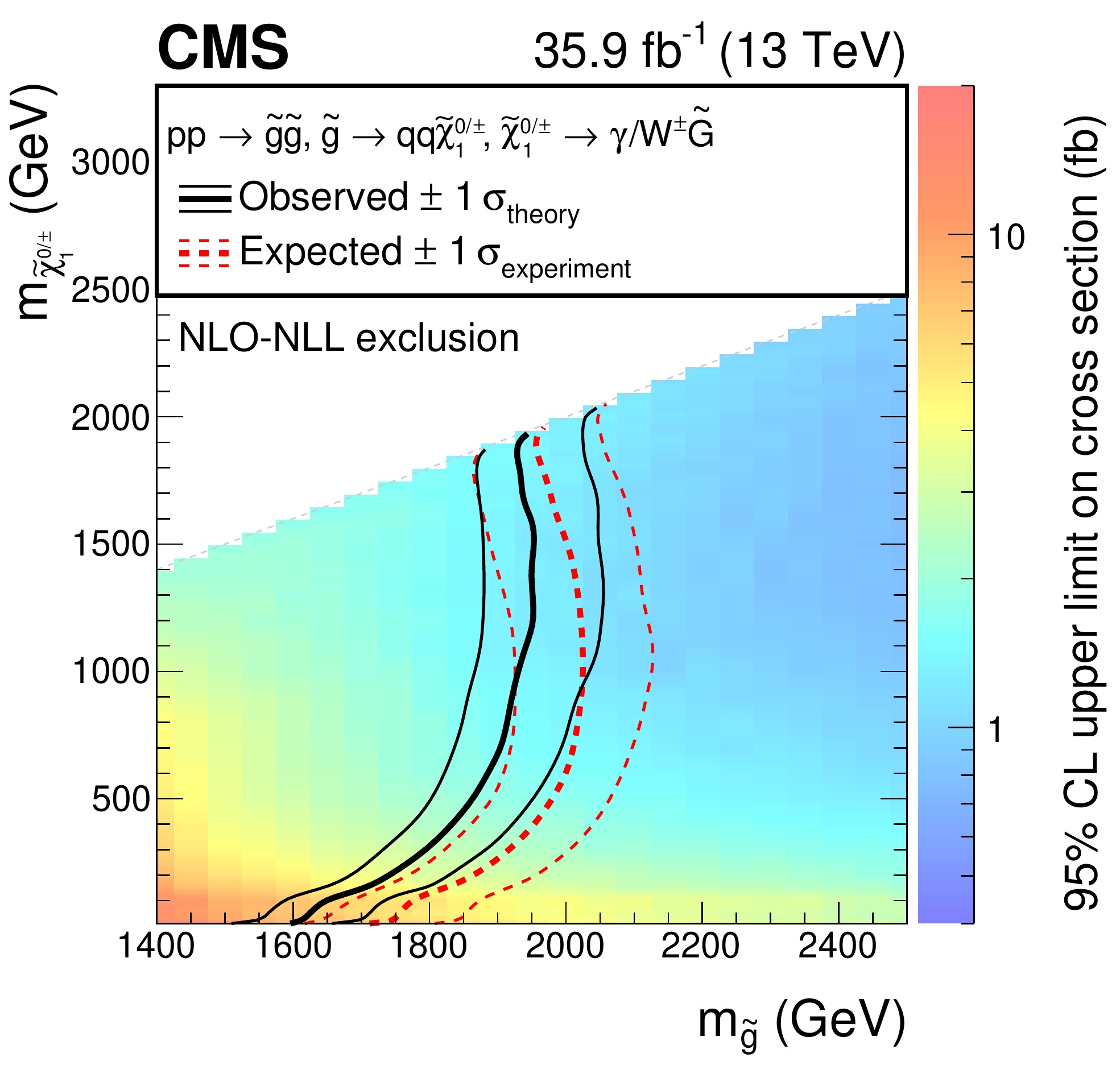}
  \caption{Exclusion limits at 95\% CL for the T6gg (top left), T6Wg (top right), T5gg (bottom left) and T5Wg (bottom right) models.
    The solid black curve represents the observed exclusion contour and the uncertainty
    due to the signal cross section.
    The red dashed curves represent the expected exclusion contours and the
    experimental uncertainties.
  }
  \label{fig:interpretation}
\end{figure}

The results are interpreted in terms of the simplified models
introduced in Section~\ref{sec:sim}.
The 95\% confidence level (CL) upper limits on the SUSY cross
section are calculated with the CL$_\text{s}$ criterion~\cite{Junk:1999kv,Read}
using the LHC-style profile likelihood ratio as test statistic~\cite{CMS-NOTE-2011-005}
evaluated in the asymptotic approximation~\cite{Cowan:2010js}.
Log-normal nuisance parameters are used to describe the systematic uncertainties.
The observed upper limits on cross sections, exclusion contours, and expected exclusion
contours are shown in Fig.~\ref{fig:interpretation}.
More stringent limits can be set on models with two photons, since the probability that
at least one photon is reconstructed is higher. In this case, for high gaugino
masses, squarks up to 1650\GeV and gluinos up to 2000\GeV can be excluded, while
for the T6Wg and T5Wg scenarios, squarks up to 1550\GeV and gluinos up to 1900\GeV
can be excluded for high gaugino masses.
The acceptance drops for low neutralino masses, since more energy is
transferred to jets, leaving less energy available for the photon and the gravitinos,
and therefore resulting in a lower value of \ptmiss.
If the chargino mass is close to the \PW~boson mass, less momentum is transferred
to the gravitino, leading to smaller \ptmiss values and, therefore, lower sensitivity.
This yields a squark mass exclusion of 1500 and 1300\GeV for the T6gg and T6Wg
model, respectively, and a gluino mass exclusion of 1750 and 1500\GeV for the T5gg
and T5Wg model, respectively.
For squark pair production, the mass exclusion is determined assuming eight
mass-degenerate squark states, corresponding to the SUSY partners of the left-
and right-handed u, d, s, and c quarks.

\section{Summary} \label{sec:summary}

A search for physics beyond the standard model (SM) in final states with at least
one photon, large missing transverse momentum, and large total transverse event activity
has been presented using data corresponding to an integrated luminosity of \reclumi of
proton--proton collisions at $\sqrt{s}=13\TeV$ recorded by the CMS experiment at the LHC in 2016.
The SM background is estimated from data and simulation, and is validated in
several control regions.
No significant signs of new physics beyond the SM are found, and the data are interpreted in
simplified models motivated by gauge-mediated supersymmetry breaking.
Gluino masses up to 1.50--2.00\TeV and squark masses up to 1.30--1.65\TeV
are excluded at 95\% confidence level, depending on the neutralino mass and mixture.

\begin{acknowledgments}
We congratulate our colleagues in the CERN accelerator departments for the excellent performance of the LHC and thank the technical and administrative staffs at CERN and at other CMS institutes for their contributions to the success of the CMS effort. In addition, we gratefully acknowledge the computing centres and personnel of the Worldwide LHC Computing Grid for delivering so effectively the computing infrastructure essential to our analyses. Finally, we acknowledge the enduring support for the construction and operation of the LHC and the CMS detector provided by the following funding agencies: BMWFW and FWF (Austria); FNRS and FWO (Belgium); CNPq, CAPES, FAPERJ, and FAPESP (Brazil); MES (Bulgaria); CERN; CAS, MoST, and NSFC (China); COLCIENCIAS (Colombia); MSES and CSF (Croatia); RPF (Cyprus); SENESCYT (Ecuador); MoER, ERC IUT, and ERDF (Estonia); Academy of Finland, MEC, and HIP (Finland); CEA and CNRS/IN2P3 (France); BMBF, DFG, and HGF (Germany); GSRT (Greece); OTKA and NIH (Hungary); DAE and DST (India); IPM (Iran); SFI (Ireland); INFN (Italy); MSIP and NRF (Republic of Korea); LAS (Lithuania); MOE and UM (Malaysia); BUAP, CINVESTAV, CONACYT, LNS, SEP, and UASLP-FAI (Mexico); MBIE (New Zealand); PAEC (Pakistan); MSHE and NSC (Poland); FCT (Portugal); JINR (Dubna); MON, RosAtom, RAS, RFBR and RAEP (Russia); MESTD (Serbia); SEIDI, CPAN, PCTI and FEDER (Spain); Swiss Funding Agencies (Switzerland); MST (Taipei); ThEPCenter, IPST, STAR, and NSTDA (Thailand); TUBITAK and TAEK (Turkey); NASU and SFFR (Ukraine); STFC (United Kingdom); DOE and NSF (USA).

{\tolerance=800
\hyphenation{Rachada-pisek} Individuals have received support from the Marie-Curie programme and the European Research Council and Horizon 2020 Grant, contract No. 675440 (European Union); the Leventis Foundation; the A. P. Sloan Foundation; the Alexander von Humboldt Foundation; the Belgian Federal Science Policy Office; the Fonds pour la Formation \`a la Recherche dans l'Industrie et dans l'Agriculture (FRIA-Belgium); the Agentschap voor Innovatie door Wetenschap en Technologie (IWT-Belgium); the Ministry of Education, Youth and Sports (MEYS) of the Czech Republic; the Council of Science and Industrial Research, India; the HOMING PLUS programme of the Foundation for Polish Science, cofinanced from European Union, Regional Development Fund, the Mobility Plus programme of the Ministry of Science and Higher Education, the National Science Center (Poland), contracts Harmonia 2014/14/M/ST2/00428, Opus 2014/13/B/ST2/02543, 2014/15/B/ST2/03998, and 2015/19/B/ST2/02861, Sonata-bis 2012/07/E/ST2/01406; the National Priorities Research Program by Qatar National Research Fund; the Programa Clar\'in-COFUND del Principado de Asturias; the Thalis and Aristeia programmes cofinanced by EU-ESF and the Greek NSRF; the Rachadapisek Sompot Fund for Postdoctoral Fellowship, Chulalongkorn University and the Chulalongkorn Academic into Its 2nd Century Project Advancement Project (Thailand); and the Welch Foundation, contract C-1845.
\par}

\end{acknowledgments}
\bibliography{auto_generated}

\providecommand{\href}[2]{#2}\begingroup\raggedright\begin{thebibliography}{10}%
\makeatletter
\providecommand{\hrefCMSnoop }[0]{\@secondoftwo}%
\makeatother
\providecommand{\doi}{\texttt{doi:}\begingroup \urlstyle{tt}\Url}

\bibitem{Barbieri198863}
\hrefCMSnoop {}{R.~Barbieri and G.~F. Giudice, ``Upper bounds on supersymmetric
  particle masses'',} \textit{ Nucl. Phys. B} \textbf{ 306} (1988) 63,
  \href{http://dx.doi.org/10.1016/0550-3213(88)90171-X}{\doi{10.1016/0550-3213(88)90171-X}}.

\bibitem{Ramond}
\hrefCMSnoop {}{P.~Ramond, ``Dual theory for free fermions'',} \textit{ Phys.
  Rev. D} \textbf{ 3} (1971) 2415,
  \href{http://dx.doi.org/10.1103/PhysRevD.3.2415}{\doi{10.1103/PhysRevD.3.2415}}.

\bibitem{Golfand:1971iw}
\href {http://www.jetpletters.ac.ru/ps/1584/article_24309.pdf}{Y.~A. Gol'fand
  and E.~P. Likhtman, ``Extension of the algebra of {P}oincar\'{e} group
  generators and violation of {P} invariance'',} \textit{ JETP Lett.} \textbf{
  13} (1971)
323.

\bibitem{Ferrara:1974pu}
\hrefCMSnoop {}{S.~Ferrara and B.~Zumino, ``Supergauge invariant yang-mills
  theories'',} \textit{ Nucl. Phys. B} \textbf{ 79} (1974) 413,
\href{http://dx.doi.org/10.1016/0550-3213(74)90559-8}{\doi{10.1016/0550-3213(74)90559-8}}.

\bibitem{Wess}
\hrefCMSnoop {}{J.~Wess and B.~Zumino, ``Supergauge transformations in
  four-dimensions'',} \textit{ Nucl. Phys. B} \textbf{ 70} (1974) 39,
  \href{http://dx.doi.org/10.1016/0550-3213(74)90355-1}{\doi{10.1016/0550-3213(74)90355-1}}.

\bibitem{Chamseddine}
\hrefCMSnoop {}{A.~H. Chamseddine, R.~L. Arnowitt, and P.~Nath, ``Locally
  supersymmetric grand unification'',} \textit{ Phys. Rev. Lett.} \textbf{ 49}
  (1982) 970,
  \href{http://dx.doi.org/10.1103/PhysRevLett.49.970}{\doi{10.1103/PhysRevLett.49.970}}.

\bibitem{Barbieri}
\hrefCMSnoop {}{R.~Barbieri, S.~Ferrara, and C.~A. Savoy, ``Gauge models with
  spontaneously broken local supersymmetry'',} \textit{ Phys. Lett. B} \textbf{
  119} (1982) 343,
  \href{http://dx.doi.org/10.1016/0370-2693(82)90685-2}{\doi{10.1016/0370-2693(82)90685-2}}.

\bibitem{Hall}
\hrefCMSnoop {}{L.~J. Hall, J.~D. Lykken, and S.~Weinberg, ``Supergravity as
  the messenger of supersymmetry breaking'',} \textit{ Phys. Rev. D} \textbf{
  27} (1983) 2359,
  \href{http://dx.doi.org/10.1103/PhysRevD.27.2359}{\doi{10.1103/PhysRevD.27.2359}}.

\bibitem{GGMa}
\hrefCMSnoop {}{P.~Fayet, ``Mixing between gravitational and weak interactions
  through the massive gravitino'',} \textit{ Phys. Lett. B} \textbf{ 70} (1977)
  461,
  \href{http://dx.doi.org/10.1016/0370-2693(77)90414-2}{\doi{10.1016/0370-2693(77)90414-2}}.

\bibitem{GGMd2}
\hrefCMSnoop {}{H.~Baer, M.~Brhlik, C.-h. Chen, and X.~Tata, ``Signals for the
  minimal gauge-mediated supersymmetry breaking model at the {Fermilab
  Tevatron} collider'',} \textit{ Phys. Rev. D} \textbf{ 55} (1997) 4463,
  \href{http://dx.doi.org/10.1103/PhysRevD.55.4463}{\doi{10.1103/PhysRevD.55.4463}},
  \href{http://www.arXiv.org/abs/hep-ph/9610358}{\texttt{arXiv:hep-ph/9610358}}.

\bibitem{GGMd3}
\hrefCMSnoop {}{H.~Baer, P.~G. Mercadante, X.~Tata, and Y.~Wang, ``Reach of
  {Fermilab Tevatron} upgrades in gauge-mediated supersymmetry breaking
  models'',} \textit{ Phys. Rev. D} \textbf{ 60} (1999) 055001,
  \href{http://dx.doi.org/10.1103/PhysRevD.60.055001}{\doi{10.1103/PhysRevD.60.055001}},
  \href{http://www.arXiv.org/abs/hep-ph/9903333}{\texttt{arXiv:hep-ph/9903333}}.

\bibitem{GGMd4}
\hrefCMSnoop {}{S.~Dimopoulos, S.~Thomas, and J.~D. Wells, ``Sparticle
  spectroscopy and electroweak symmetry breaking with gauge-mediated
  supersymmetry breaking'',} \textit{ Nucl. Phys. B} \textbf{ 488} (1997) 39,
  \href{http://dx.doi.org/10.1016/S0550-3213(97)00030-8}{\doi{10.1016/S0550-3213(97)00030-8}},
  \href{http://www.arXiv.org/abs/hep-ph/9609434}{\texttt{arXiv:hep-ph/9609434}}.

\bibitem{GGMd5}
\hrefCMSnoop {}{J.~Ellis, J.~L. Lopez, and D.~V. Nanopoulos, ``Analysis of
  {LEP} constraints on supersymmetric models with a light gravitino'',}
  \textit{ Phys. Lett. B} \textbf{ 394} (1997) 354,
  \href{http://dx.doi.org/10.1016/S0370-2693(97)00019-1}{\doi{10.1016/S0370-2693(97)00019-1}},
  \href{http://www.arXiv.org/abs/hep-ph/9610470}{\texttt{arXiv:hep-ph/9610470}}.

\bibitem{GGMd1}
\hrefCMSnoop {}{M.~Dine, A.~E. Nelson, Y.~Nir, and Y.~Shirman, ``New tools for
  low energy dynamical supersymmetry breaking'',} \textit{ Phys. Rev. D}
  \textbf{ 53} (1996) 2658,
  \href{http://dx.doi.org/10.1103/PhysRevD.53.2658}{\doi{10.1103/PhysRevD.53.2658}},
  \href{http://www.arXiv.org/abs/hep-ph/9507378}{\texttt{arXiv:hep-ph/9507378}}.

\bibitem{GGMd}
\hrefCMSnoop {}{G.~F. Giudice and R.~Rattazzi, ``Gauge-mediated supersymmetry
  breaking'',} in \textit{ Perspectives on Supersymmetry}, p.~355.
\newblock World Scientific, Singapore, 1998.

\bibitem{Grajek:2013ola}
\hrefCMSnoop {}{P.~Grajek, A.~Mariotti, and D.~Redigolo, ``{Phenomenology of
  general gauge mediation in light of a 125\GeV Higgs}'',} \textit{ JHEP}
  \textbf{ 07} (2013) 109,
  \href{http://dx.doi.org/10.1007/JHEP07(2013)109}{\doi{10.1007/JHEP07(2013)109}},
\href{http://www.arXiv.org/abs/1303.0870}{\texttt{arXiv:1303.0870}}.

\bibitem{rparity}
\hrefCMSnoop {}{G.~R. Farrar and P.~Fayet, ``Phenomenology of the production,
  decay, and detection of new hadronic states associated with supersymmetry'',}
  \textit{ Phys. Lett. B} \textbf{ 76} (1978) 575,
  \href{http://dx.doi.org/10.1016/0370-2693(78)90858-4}{\doi{10.1016/0370-2693(78)90858-4}}.

\bibitem{Barbier:2004ez}
R.~Barbier\hrefCMSnoop {}{ {et~al.}, ``{$R$}-parity-violating supersymmetry'',}
  \textit{ Phys. Rep.} \textbf{ 420} (2005) 1,
  \href{http://dx.doi.org/10.1016/j.physrep.2005.08.006}{\doi{10.1016/j.physrep.2005.08.006}},
  \href{http://www.arXiv.org/abs/hep-ph/0406039}{\texttt{arXiv:hep-ph/0406039}}.

\bibitem{Ruderman:2011vv}
\hrefCMSnoop {}{J.~T. Ruderman and D.~Shih, ``General neutralino {NLSP}s at the
  early {LHC}'',} \textit{ JHEP} \textbf{ 08} (2012) 159,
  \href{http://dx.doi.org/10.1007/JHEP08(2012)159}{\doi{10.1007/JHEP08(2012)159}},
  \href{http://www.arXiv.org/abs/1103.6083}{\texttt{arXiv:1103.6083}}.

\bibitem{CMS-PAPERS-SUS-14-004}
\hrefCMSnoop {}{{CMS Collaboration}, ``Search for supersymmetry with photons in
  pp collisions at {$\sqrt{s}=8\TeV$}'',} \textit{ Phys. Rev. D} \textbf{ 92}
  (2015) 072006,
  \href{http://dx.doi.org/10.1103/PhysRevD.92.072006}{\doi{10.1103/PhysRevD.92.072006}},
  \href{http://www.arXiv.org/abs/1507.02898}{\texttt{arXiv:1507.02898}}.

\bibitem{Khachatryan:2016hns}
\hrefCMSnoop {}{{CMS Collaboration}, ``{Search for supersymmetry in electroweak
  production with photons and large missing transverse energy in pp collisions
  at $\sqrt{s} = 8\TeV$}'',} \textit{ Phys. Lett. B} \textbf{ 759} (2016) 479,
  \href{http://dx.doi.org/10.1016/j.physletb.2016.05.088}{\doi{10.1016/j.physletb.2016.05.088}},
\href{http://www.arXiv.org/abs/1602.08772}{\texttt{arXiv:1602.08772}}.

\bibitem{Khachatryan:2016ojf}
\hrefCMSnoop {}{{CMS Collaboration}, ``{Search for supersymmetry in events with
  photons and missing transverse energy in pp collisions at 13 TeV}'',}
  \textit{ Phys. Lett. B} \textbf{ 769} (2017) 391,
  \href{http://dx.doi.org/10.1016/j.physletb.2017.04.005}{\doi{10.1016/j.physletb.2017.04.005}},
\href{http://www.arXiv.org/abs/1611.06604}{\texttt{arXiv:1611.06604}}.

\bibitem{Aad:2015hea}
\hrefCMSnoop {}{{ATLAS Collaboration}, ``{Search for photonic signatures of
  gauge-mediated supersymmetry in 8 TeV pp collisions with the ATLAS
  detector}'',} \textit{ Phys. Rev. D} \textbf{ 92} (2015) 072001,
  \href{http://dx.doi.org/10.1103/PhysRevD.92.072001}{\doi{10.1103/PhysRevD.92.072001}},
\href{http://www.arXiv.org/abs/1507.05493}{\texttt{arXiv:1507.05493}}.

\bibitem{ATLASCollaboration:2016wlb}
\hrefCMSnoop {}{{ATLAS Collaboration}, ``{Search for supersymmetry in a final
  state containing two photons and missing transverse momentum in $\sqrt{s} =
  13 \TeV$ pp collisions at the LHC using the ATLAS detector}'',} \textit{ Eur.
  Phys. J. C} \textbf{ 76} (2016) 517,
  \href{http://dx.doi.org/10.1140/epjc/s10052-016-4344-x}{\doi{10.1140/epjc/s10052-016-4344-x}},
\href{http://www.arXiv.org/abs/1606.09150}{\texttt{arXiv:1606.09150}}.

\bibitem{Khachatryan:2016kdb}
\hrefCMSnoop {}{{CMS Collaboration}, ``{Jet energy scale and resolution in the
  CMS experiment in pp collisions at 8 TeV}'',} \textit{ JINST} \textbf{ 12}
  (2017) P02014,
  \href{http://dx.doi.org/10.1088/1748-0221/12/02/P02014}{\doi{10.1088/1748-0221/12/02/P02014}},
\href{http://www.arXiv.org/abs/1607.03663}{\texttt{arXiv:1607.03663}}.

\bibitem{Chatrchyan:2008zzk}
\hrefCMSnoop {}{{CMS Collaboration}, ``The {CMS} experiment at the {CERN}
  {LHC}'',} \textit{ JINST} \textbf{ 3} (2008) S08004,
  \href{http://dx.doi.org/10.1088/1748-0221/3/08/S08004}{\doi{10.1088/1748-0221/3/08/S08004}}.

\bibitem{Sirunyan:2017ulk}
\hrefCMSnoop {}{{CMS Collaboration}, ``{Particle-flow reconstruction and global
  event description with the CMS detector}'',} (2017).
  \href{http://www.arXiv.org/abs/1706.04965}{\texttt{arXiv:1706.04965}}.
Submitted to \textit{JINST}.

\bibitem{Cacciari:2008gp}
\hrefCMSnoop {}{M.~Cacciari, G.~P. Salam, and G.~Soyez, ``The anti-$k_t$ jet
  clustering algorithm'',} \textit{ JHEP} \textbf{ 04} (2008) 063,
  \href{http://dx.doi.org/10.1088/1126-6708/2008/04/063}{\doi{10.1088/1126-6708/2008/04/063}},
  \href{http://www.arXiv.org/abs/0802.1189}{\texttt{arXiv:0802.1189}}.

\bibitem{Cacciari:2011ma}
\hrefCMSnoop {}{M.~Cacciari, G.~P. Salam, and G.~Soyez, ``Fastjet user
  manual'',} \textit{ Eur. Phys. J. C} \textbf{ 72} (2012) 1896,
  \href{http://dx.doi.org/10.1140/epjc/s10052-012-1896-2}{\doi{10.1140/epjc/s10052-012-1896-2}},
\href{http://www.arXiv.org/abs/1111.6097}{\texttt{arXiv:1111.6097}}.

\bibitem{Cacciari:2007fd}
\hrefCMSnoop {}{M.~Cacciari and G.~P. Salam, ``Pileup subtraction using jet
  areas'',} \textit{ Phys. Lett. B} \textbf{ 659} (2008) 119,
  \href{http://dx.doi.org/10.1016/j.physletb.2007.09.077}{\doi{10.1016/j.physletb.2007.09.077}},
\href{http://www.arXiv.org/abs/0707.1378}{\texttt{arXiv:0707.1378}}.

\bibitem{Alwall:2014hca}
J.~Alwall\hrefCMSnoop {}{ {et~al.}, ``{The automated computation of tree-level
  and next-to-leading order differential cross sections, and their matching to
  parton shower simulations}'',} \textit{ JHEP} \textbf{ 07} (2014) 079,
  \href{http://dx.doi.org/10.1007/JHEP07(2014)079}{\doi{10.1007/JHEP07(2014)079}},
\href{http://www.arXiv.org/abs/1405.0301}{\texttt{arXiv:1405.0301}}.

\bibitem{Ball:2014uwa}
\hrefCMSnoop {}{{NNPDF} Collaboration, ``{Parton distributions for the LHC Run
  II}'',} \textit{ JHEP} \textbf{ 04} (2015) 040,
  \href{http://dx.doi.org/10.1007/JHEP04(2015)040}{\doi{10.1007/JHEP04(2015)040}},
\href{http://www.arXiv.org/abs/1410.8849}{\texttt{arXiv:1410.8849}}.

\bibitem{Sjostrand:2007gs}
\hrefCMSnoop {}{T.~Sj{\"o}strand, S.~Mrenna, and P.~Z. Skands, ``A brief
  introduction to {PYTHIA 8.1}'',} \textit{ Comput. Phys. Commun.} \textbf{
  178} (2008) 852,
  \href{http://dx.doi.org/10.1016/j.cpc.2008.01.036}{\doi{10.1016/j.cpc.2008.01.036}},
\href{http://www.arXiv.org/abs/0710.3820}{\texttt{arXiv:0710.3820}}.

\bibitem{Khachatryan:2015pea}
\hrefCMSnoop {}{{CMS Collaboration}, ``{Event generator tunes obtained from
  underlying event and multiparton scattering measurements}'',} \textit{ Eur.
  Phys. J. C} \textbf{ 76} (2016) 155,
  \href{http://dx.doi.org/10.1140/epjc/s10052-016-3988-x}{\doi{10.1140/epjc/s10052-016-3988-x}},
\href{http://www.arXiv.org/abs/1512.00815}{\texttt{arXiv:1512.00815}}.

\bibitem{Borschensky:2014cia}
C.~Borschensky\hrefCMSnoop {}{ {et~al.}, ``Squark and gluino production cross
  sections in pp collisions at $\sqrt{s} = 13, 14, 33$ and {$100\TeV$}'',}
  \textit{ Eur. Phys. J. C} \textbf{ 74} (2014) 3174,
  \href{http://dx.doi.org/10.1140/epjc/s10052-014-3174-y}{\doi{10.1140/epjc/s10052-014-3174-y}},
\href{http://www.arXiv.org/abs/1407.5066}{\texttt{arXiv:1407.5066}}.

\bibitem{Alves:2011wf}
\hrefCMSnoop {}{{LHC New Physics Working Group}, ``Simplified models for {LHC}
  new physics searches'',} \textit{ J. Phys. G} \textbf{ 39} (2012) 105005,
  \href{http://dx.doi.org/10.1088/0954-3899/39/10/105005}{\doi{10.1088/0954-3899/39/10/105005}},
  \href{http://www.arXiv.org/abs/1105.2838}{\texttt{arXiv:1105.2838}}.

\bibitem{Chatrchyan:2013sza}
\hrefCMSnoop {}{{CMS Collaboration}, ``{Interpretation of searches for
  supersymmetry with simplified models}'',} \textit{ Phys. Rev. D} \textbf{ 88}
  (2013) 052017,
  \href{http://dx.doi.org/10.1103/PhysRevD.88.052017}{\doi{10.1103/PhysRevD.88.052017}},
\href{http://www.arXiv.org/abs/1301.2175}{\texttt{arXiv:1301.2175}}.

\bibitem{Agostinelli:2002hh}
\hrefCMSnoop {}{{GEANT4} Collaboration, ``{GEANT4}---a simulation toolkit'',}
  \textit{ Nucl. Instrum. Meth. A} \textbf{ 506} (2003) 250,
\href{http://dx.doi.org/10.1016/S0168-9002(03)01368-8}{\doi{10.1016/S0168-9002(03)01368-8}}.

\bibitem{CMS-DP-2010-039}
\hrefCMSnoop {}{{CMS Collaboration}, ``The fast simulation of the {CMS}
  detector at {LHC}'',} in \textit{ {Int'l. Conf. on High Energy and Nuclear
  Physics}}.
\newblock Taipei, Taiwan, October, 2011.
\newblock [{J. Phys. Conf. Ser. 331 (2011) 032049}].
  \href{http://dx.doi.org/10.1088/1742-6596/331/3/032049}{\doi{10.1088/1742-6596/331/3/032049}}.

\bibitem{Sekmen:2017hzs}
\href {https://pos.sissa.it/archive/conferences/282/181/ICHEP2016_181.pdf}{{CMS
  Collaboration}, ``Recent developments in {CMS} fast simulation'',} in
  \textit{ Proceedings, 38th International Conference on High Energy Physics
  ({ICHEP 2016})}, p.~181.
\newblock Chicago, Illinois, USA, August, 2016.
\newblock
  \href{http://www.arXiv.org/abs/1701.03850}{\texttt{arXiv:1701.03850}}.
\newblock
{[PoS(ICHEP2016)181}].

\bibitem{Khachatryan:2016bia}
\hrefCMSnoop {}{{CMS Collaboration}, ``{The CMS trigger system}'',} \textit{
  JINST} \textbf{ 12} (2017) P01020,
  \href{http://dx.doi.org/10.1088/1748-0221/12/01/P01020}{\doi{10.1088/1748-0221/12/01/P01020}},
\href{http://www.arXiv.org/abs/1609.02366}{\texttt{arXiv:1609.02366}}.

\bibitem{CMS-PAS-JME-16-004}
\href {http://cdsweb.cern.ch/record/2205284}{{CMS Collaboration}, ``Performance
  of missing energy reconstruction in 13 {TeV} pp collision data using the
  {CMS} detector'',} CMS Physics Analysis Summary CMS-PAS-JME-16-004, 2016.

\bibitem{Butterworth:2015oua}
\hrefCMSnoop {}{J.~Butterworth {et~al.}, ``{PDF4LHC} recommendations for {LHC
  Run II}'',} \textit{ J. Phys. G} \textbf{ 43} (2016) 023001,
  \href{http://dx.doi.org/10.1088/0954-3899/43/2/023001}{\doi{10.1088/0954-3899/43/2/023001}},
\href{http://www.arXiv.org/abs/1510.03865}{\texttt{arXiv:1510.03865}}.

\bibitem{CMS-PAS-LUM-17-001}
\href {http://cdsweb.cern.ch/record/2257069}{{CMS Collaboration}, ``{CMS}
  luminosity measurements for the 2016 data taking period'',} CMS Physics
  Analysis Summary CMS-PAS-LUM-17-001, 2017.

\bibitem{Junk:1999kv}
\hrefCMSnoop {}{T.~Junk, ``{Confidence level computation for combining searches
  with small statistics}'',} \textit{ Nucl. Instrum. Meth. A} \textbf{ 434}
  (1999) 435,
  \href{http://dx.doi.org/10.1016/S0168-9002(99)00498-2}{\doi{10.1016/S0168-9002(99)00498-2}},
\href{http://www.arXiv.org/abs/hep-ex/9902006}{\texttt{arXiv:hep-ex/9902006}}.

\bibitem{Read}
\hrefCMSnoop {}{A.~L. Read, ``Presentation of search results: the {$CL_s$}
  technique'',} \textit{ J. Phys. G} \textbf{ 28} (2002) 2693,
\href{http://dx.doi.org/10.1088/0954-3899/28/10/313}{\doi{10.1088/0954-3899/28/10/313}}.

\bibitem{CMS-NOTE-2011-005}
\href {http://cdsweb.cern.ch/record/1379837}{{ATLAS and CMS Collaborations},
  ``Procedure for the {LHC Higgs} boson search combination in {Summer 2011}'',}
  Technical Report CMS-NOTE-2011-005, ATL-PHYS-PUB-2011-11, 2011.

\bibitem{Cowan:2010js}
\hrefCMSnoop {}{G.~Cowan, K.~Cranmer, E.~Gross, and O.~Vitells, ``Asymptotic
  formulae for likelihood-based tests of new physics'',} \textit{ Eur. Phys. J.
  C} \textbf{ 71} (2011) 1554,
  \href{http://dx.doi.org/10.1140/epjc/s10052-011-1554-0}{\doi{10.1140/epjc/s10052-011-1554-0}},
  \href{http://www.arXiv.org/abs/1007.1727}{\texttt{arXiv:1007.1727}}.
[Erratum: \DOI{10.1140/epjc/s10052-013-2501-z}].

\end{thebibliography}\endgroup

\cleardoublepage \appendix\section{The CMS Collaboration \label{app:collab}}\begin{sloppypar}\hyphenpenalty=5000\widowpenalty=500\clubpenalty=5000\textbf{Yerevan Physics Institute,  Yerevan,  Armenia}\\*[0pt]
A.M.~Sirunyan, A.~Tumasyan
\vskip\cmsinstskip
\textbf{Institut f\"{u}r Hochenergiephysik,  Wien,  Austria}\\*[0pt]
W.~Adam, F.~Ambrogi, E.~Asilar, T.~Bergauer, J.~Brandstetter, E.~Brondolin, M.~Dragicevic, J.~Er\"{o}, M.~Flechl, M.~Friedl, R.~Fr\"{u}hwirth\cmsAuthorMark{1}, V.M.~Ghete, J.~Grossmann, J.~Hrubec, M.~Jeitler\cmsAuthorMark{1}, A.~K\"{o}nig, N.~Krammer, I.~Kr\"{a}tschmer, D.~Liko, T.~Madlener, I.~Mikulec, E.~Pree, D.~Rabady, N.~Rad, H.~Rohringer, J.~Schieck\cmsAuthorMark{1}, R.~Sch\"{o}fbeck, M.~Spanring, D.~Spitzbart, J.~Strauss, W.~Waltenberger, J.~Wittmann, C.-E.~Wulz\cmsAuthorMark{1}, M.~Zarucki
% ...
\vskip\cmsinstskip
% ...
1:~~Also at Vienna University of Technology, Vienna, Austria\\
% ...

\end{sloppypar}
\end{document}